\documentclass[times,twocolumn,final]{elsarticle}

\usepackage[title,titletoc]{appendix}
\usepackage{yjvci}
\usepackage{framed,multirow}

\usepackage{amssymb}
\usepackage{latexsym}

\usepackage{url}

\usepackage{hyperref}
\usepackage{amsmath,amssymb}
\usepackage{chngpage}
\usepackage{bbm}
\usepackage{makecell}
\usepackage[table,xcdraw]{xcolor}
\usepackage{tablefootnote}
\usepackage{tabularx}

\definecolor{newcolor}{rgb}{.8,.349,.1}

\journal{Journal of Visual Communication and Image Representation}

\begin{document}

\verso{Alex Golts \textit{et~al.}}

\begin{frontmatter}


\title{Image compression optimized for 3D reconstruction by utilizing deep neural networks}%

\author[1]{Alex \snm{Golts}\corref{cor1}}
\cortext[cor1]{Corresponding author:}
\ead{alex@golts.net}
\author[2]{Yoav Y. \snm{Schechner}}

\address[1]{Rafael Advanced Defense Systems LTD., Israel, e-mail: alex@golts.net}
\address[2]{Viterbi Faculty of Electrical Engineering, Technion - Israel Institute of Technology, Haifa, Israel. e-mail: yoav@ee.technion.ac.il}


\begin{abstract}
Computer vision tasks are often expected to be executed on compressed images. Classical image compression standards like JPEG 2000 are widely used. However, they do not account for the specific end-task at hand. Motivated by works on recurrent neural network (RNN)-based image compression and three-dimensional (3D) reconstruction, we propose unified network architectures to solve both tasks jointly. These joint models provide image compression tailored for the specific task of 3D reconstruction. Images compressed by our proposed models, yield 3D reconstruction performance superior as compared to using JPEG 2000 compression. Our models significantly extend the range of compression rates for which 3D reconstruction is possible. We also show that this can be done highly efficiently at almost no additional cost to obtain compression on top of the computation already required for performing the 3D reconstruction task.
\end{abstract}


\end{frontmatter}


\section{Introduction}
\label{introduction}
Image compression is an essential step in many image processing and computer vision pipelines. At times, the sole goal of compression is to deliver images that a human viewer would perceive as having high quality given some compression ratio.
However, computer vision systems are often autonomous. This means that evaluation criteria and compression goals should be 
tailored to computer vision tasks. This paper focuses on the computer vision task of multi-view three-dimensional (3D) reconstruction. Compression is thus jointly optimized with 3D reconstruction.

Compression in this context is important when multi-view data has to be transmitted using limited resources. Specifically, this applies to imaging using drones (\cite{renwick2016drone}), airplanes (\cite{vetrivel2015identification}) and satellites (\cite{facciolo2017automatic}), where power and connectivity are limited. For 3D reconstruction, computation is applied on the transmitted images to extract 3D information in the form of volumetric occupancy grids (\cite{choy20163d},~\cite{xie2019pix2vox}), point clouds, depth maps or surface mesh models (\cite{fuhrmann2014mve}). The images are typically compressed  (\cite{indradjad2019comparison}), however, the compression methods that are mostly used are well known standards tailored to image quality metrics, and not directly to evaluation metrics of 3D reconstruction.     

To solve our 
problem we use a deep neural network (DNN). DNN-based methods are increasingly successful and popular at solving various image processing and computer vision tasks.
In particular, \cite{toderici2017full} proposed a recurrent neural network (RNN)-based method for image compression that outperforms the well-known JPEG standard~\cite{wallace1992jpeg}.
More recently, several non-RNN based compression methods were proposed. They outperformed \cite{toderici2017full} achieving compression performance competitive \cite{theis2017lossy}, or better than (\cite{balle2016end}, \cite{balle2018variational}) the JPEG 2000 standard~\cite{rabbani2002overview}. Some \cite{minnen2018joint} even exceeding the performance of the more recent BPG codec~\cite{bellard_bpg}. Neural networks are a natural choice for image compression, as they often compress their input signal, even when compression is not their explicit goal.
For the task of stereoscopic 3D reconstruction, \cite{choy20163d}  proposed an RNN architecture that learns a mapping from images of objects at different viewpoints, to a 3D occupancy grid corresponding to the object's shape. 
More recently, \cite{xie2019pix2vox} improved 3D reconstruction performance while using a more computationally efficient non RNN-based approach.
When applying image compression to multiple view scenarios where a large overlap exists between adjacent images, some works~(\cite{liu2019dsic}) leverage the redundancy in the overlapped regions to compress better.  

We postulate that jointly learning these two tasks (image compression and 3D reconstruction) may lead to compression better suited for 3D reconstruction. 
\cite{zamir2018taskonomy} showed that learning multiple visual tasks jointly requires less labeled data to achieve the performance obtained by separate learning-based systems. However, this was shown for small auxiliary sub-tasks that aid in a more complex grand task. In our case, both compression and 3D recovery are critical components. Moreover, one can hardly expect compression to actually \textit{help} the mission. It is a necessity. As compared to stand-alone compression, it can be expected that tailored compression enhances the mission performance, for a given compression rate. 

In the context of image compression, \cite{torfason2018towards} showed that image understanding tasks such as classification and semantic segmentation can be performed directly on compressed representations derived by DNNs . This alleviates the need to decompress files prior to image understanding. They also further show that by jointly training image compression and classification networks, synergies are obtained leading to performance gains in both tasks.

In this work we propose methods to compress images so that they can be optimally used for 3D reconstruction. Our method can work with only negligible computations for image compression, on top of a system for 3D reconstruction based on \cite{choy20163d}. It also exceeds the 3D reconstruction performance obtained from images compressed by JPEG-2000~\cite{rabbani2002overview} or learned compression, across a wide range of medium to ultra aggressive compression rates that we focused on. 
Our main focus in this work is on the regime of high compression rates. We find that lower compression rates may be better suited when the goal is to obtain visually satisfying decompressed images. When the ultimate task is different, it is possible to compress images further. We study these limits for the task of 3D reconstruction.

Our approach uses RNN-based basic components for image compression \cite{toderici2017full} and 3D reconstruction \cite{choy20163d}. It does not aim to propose the most current state-of-the-art for these separate tasks. Rather, we wish to illustrate a generic concept involving the multi-task learning of these tasks. We demonstrate the benefits in considering such a framework, while assuming the above RNN-based building blocks. We hope that our findings provides motivation for further research around this subject, which may also be tailored for using more current basic methods.

\section{Background}
\label{background}
Sec.~\ref{sec:background_rnns} provides basic background on RNNs and LSTMs. The reader familiar with these topics may skip to Secs.~\ref{sec:background_compression},~\ref{sec:background_3d_reconstruction}, which provide background on specific works in RNN-based image compression and 3D reconstruction that we build on.
\subsection{Recurrent neural networks}
\label{sec:background_rnns}
\subsubsection{Motivation}
A feedforward neural network's forward pass accepts a single fixed size input vector $\mathbf{x}$. The network applies a fixed amount of computations, depending on the network depth and architecture. The network then outputs a single fixed size output vector $\mathbf{y}$. A common example is image classification, where, $\mathbf{x}$ is an image and $\mathbf{y}$ is a vector of probabilities that the imaged object belongs to a class. 
Consider for a moment $\mathbf{x}= \left\{ \mathbf{x}_{t} \right\}$ to be a movie sequence, where $t$ denotes a time step. The task is to classify events over time (\cite{colah}). Proper understanding of a moment in the movie should rely on understanding of previous frames. Feedforward neural networks cannot reason about previous events to inform later ones. RNNs address this issue.

\begin{figure*}
  \centering
    \includegraphics[width=0.8\textwidth]{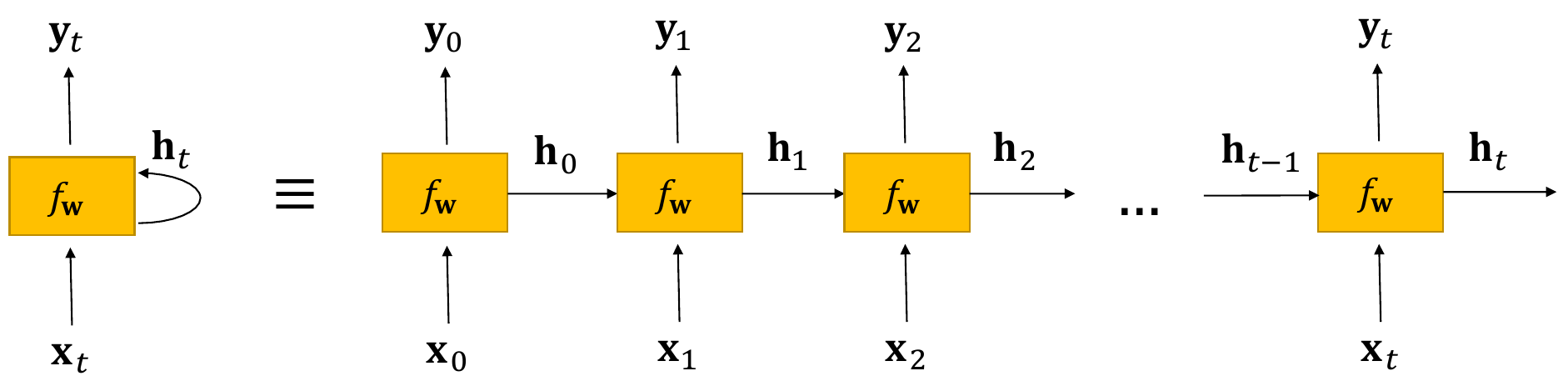}
     \caption{Block diagram of an RNN in its cycled (left) and equivalent unrolled form (right)}
\label{fig:rnn}
\end{figure*}

Fig.~\ref{fig:rnn} shows a block diagram of an RNN in its cycled (left), and equivalent unrolled form (right). 
Here $f_{\mathbf{w}}$ is a neural network block with parameters $\bf{w}$, which is a recurrence formula applied at every time step $t$ on the current input $\mathbf{x}_{t}$. A loop shown in Fig.~\ref{fig:rnn} (left) allows information to be passed from one time step of the network to the next. The unrolled form (right) reveals that an RNN can be thought of as simply multiple copies of the same network layer, each passing a message to a successor.
The recurrence formula $f_{\mathbf{w}}$ in its general form is 
\begin{equation}
\label{eq:general_recurrence_formula}
\left\{ \mathbf{h}_{t}, \mathbf{y}_{t} \right\} =f_{\mathbf{w}}\left(\mathbf{h}_{t-1},\mathbf{x}_{t}\right)
\end{equation}
Eq.~(\ref{eq:general_recurrence_formula}) operates on the current input $\mathbf{x}_{t}$ and the RNN's hidden state $\mathbf{h}_{t-1}$ from the previous time step. The outputs of  Eq.~(\ref{eq:general_recurrence_formula}) are $\mathbf{h}_{t}$ and $\mathbf{y}_{t}$, the RNN's hidden state and output in the current time step, respectively.
Contrary to feedforward neural networks, RNNs commonly operate over \textit{sequences} of vectors, rather than an individual vector (\cite{karpathy_rnns}). 
\subsubsection{Sequential processing of individual vectors}
It is sometimes beneficial to apply RNNs even when both the input and output are individual, fixed size vectors (See \cite{shyam2017attentive}).
For example, in image compression, (Sec.~\ref{sec:background_compression}), the input is a fixed size single image, and the output is either a binary compressed code, or a fixed size image. 
An output compressed code is often of a variable length, depending on the desired compression rate.
While such compression \textit{can} be performed using a feedforward neural network, it can benefit from the sequential processing power and memory capability of RNNs (\cite{toderici2015variable}, \cite{toderici2017full}). 
\subsubsection{Vanilla RNNs}
In a standard (vanilla) RNN, the block $f_{\bf{w}}$ performs a simple recurring operation
\begin{equation}
\label{eq:vanilla_rnn}
{\mathbf{h}}_{t} = \mathrm{tanh}\left(\mathbf{W}_{\mathrm{h}}{{\mathbf{h}}}_{t-1}+\mathbf{W}_{\mathrm{x}}{\mathbf{x}}_{t}\right),~~~~
{\mathbf{y}}_{t} = \mathbf{W}_{\mathrm{y}} {\mathbf{h}}_{t}.
\end{equation}
Here the $\mathrm{tanh}$ activation function is applied elementwise, and ${\bf{W}}_{h}$, ${\bf{W}}_{x}$ and ${\bf{W}}_{y}$ are learned parameters (weights) of the RNN. 
Multiple such RNN layers can be concatenated to form deep RNNs.
In practice, such standard RNNs suffer from difficulty in learning long-term dependencies throughout a sequence (\cite{bengio1994learning}).
Therefore, a special kind of RNN called Long-Short Term Memory (LSTM) network, was proposed by \cite{hochreiter1997long}, and adopted widely and successfully.
\subsubsection{LSTM networks}     
\begin{figure*}
  \centering
    \includegraphics[width=0.8\textwidth]{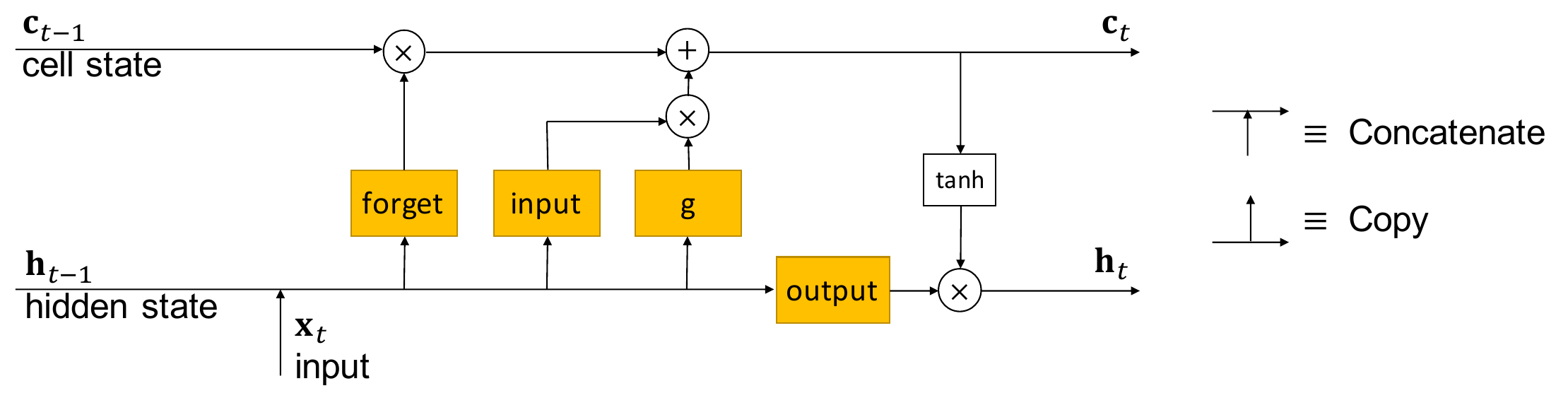}
     \caption{Block diagram of an LSTM. A notation legend appears on the right.}
\label{fig:lstm}
\end{figure*}

LSTMs are explicitly designed to avoid the long-term dependency problem.
When $f_{\bf{w}}$ is an LSTM, it has a more complicated structure compared to the vanilla RNN. Its block diagram is shown in Fig.~\ref{fig:lstm}.
Mathematically, the LSTM recurrence formulae are given by a set of equations as follows. In these equations, $\odot$ denotes element-wise multiplication, $\sigma$ denotes the Sigmoid function $\sigma(u) = \left[ 1+e^{-u} \right]^{-1} $, and $\mathbf{b}_{f}$, $\mathbf{b}_{i}$, $\mathbf{b}_{o}$ and $\mathbf{b}_{c}$ are learned bias vectors.

There are three basic \textit{gates} involved in LSTM computation: \textit{forget}, \textit{input} and \textit{output} gates, which are respectively:
\begin{equation}
\label{eq:lstm_forget_gate}
\mathbf{f}_{t} = \sigma\left(\mathbf{W}_{\mathrm{fx}}\mathbf{x}_{t}+\mathbf{W}_{\mathrm{fh}}\mathbf{h}_{t-1}+\mathbf{b}_{\mathrm{f}}\right),
\end{equation}
\begin{equation}
\label{eq:lstm_input_gate}
\mathbf{i}_{t} = \sigma\left(\mathbf{W}_{\mathrm{ix}}\mathbf{x}_{t}+\mathbf{W}_{\mathrm{ih}}\mathbf{h}_{t-1}+\mathbf{b}_{\mathrm{i}}\right),
\end{equation}
\begin{equation}
\label{eq:lstm_output_gate}
\mathbf{o}_{t} = \sigma\left(\mathbf{W}_{\mathrm{ox}}\mathbf{x}_{t}+\mathbf{W}_{\mathrm{oh}}\mathbf{h}_{t-1}+\mathbf{b}_{\mathrm{o}}\right).
\end{equation}
Another gate, which usually does not have an explicit name in the literature, is
\begin{equation}
\label{eq:lstm_g_gate}
\mathbf{g}_{t} = \mathrm{tanh}\left(\mathbf{W}_{\mathrm{cx}}\mathbf{x}_{t}+\mathbf{W}_{\mathrm{ch}}\mathbf{h}_{t-1}+\mathbf{b}_{\mathrm{c}}\right),
\end{equation}
The \textit{cell state} of the LSTM, is defined by 
\begin{equation}
\label{eq:lstm_cell_state}
\mathbf{c}_{t} = \mathbf{f}_{t} \odot \mathbf{c}_{t-1} + \mathbf{i}_{t} \odot \mathbf{g}_{t},
\end{equation}
The hidden state $\mathbf{h}_{t}$ depends on the cell state through
\begin{equation}
\label{eq:lstm_hidden_state}
\mathbf{h}_{t} = \mathbf{o}_{t} \odot \mathrm{tanh}\left( \mathbf{c}_{t} \right),
\end{equation}
In the case of an LSTM, the hidden state is also the final RNN output (Eqs.~(\ref{eq:general_recurrence_formula}-\ref{eq:vanilla_rnn}))
\begin{equation}
\label{eq:lstm_output}
\mathbf{y}_{t} \equiv \mathbf{h}_{t}.
\end{equation}
Intuitively, and following Eq.~(\ref{eq:lstm_cell_state}), the forget gate regulates whether to forget the previous cell state $\mathbf{c}_{t-1}$ in calculating the current cell state $\mathbf{c}_{t}$. The input gate regulates whether to write the new input $\mathbf{x}_{t}$ and hidden state $\mathbf{h}_{t}$ to the current cell state. The gate $\mathbf{g}_{t}$ weights \textit{how much} of these new vectors to write to the cell state. Finally, the output gate regulates how much to reveal the current cell state for updating the hidden state.

This formulation shows that an LSTM unit explicitly controls the flow from input to output. The gates regulate how much and which information from the previous time steps and current input, proceeds to the next time step. 
It can be shown~(\cite{bayer2015learning}, \cite{pascanu2013difficulty}) that in an LSTM, unlike a vanilla RNN, the gradient of the cell state with respect to itself at some previous time step, does not decay exponentially in the time step difference. This is what allows the LSTM to mitigate the vanishing or exploding gradient issue, which exists in vanilla RNNs and may prevent learning of long term dependencies.  

Multiple slight alternative formulations of the LSTM equations have been proposed. A more conceptual alternative is the Convolutional LSTM proposed by \cite{xingjian2015convolutional}, in which the matrix multiplications in Eqs.~(\ref{eq:lstm_forget_gate}-\ref{eq:lstm_g_gate}) are replaced with convolutional filtering, in analogy to convolutional neural networks (CNNs). This is especially useful for image-related tasks.  
\subsection{RNN-based image compression}
\label{sec:background_compression}
\begin{figure*}
  \centering
    \includegraphics[width=0.8\textwidth]{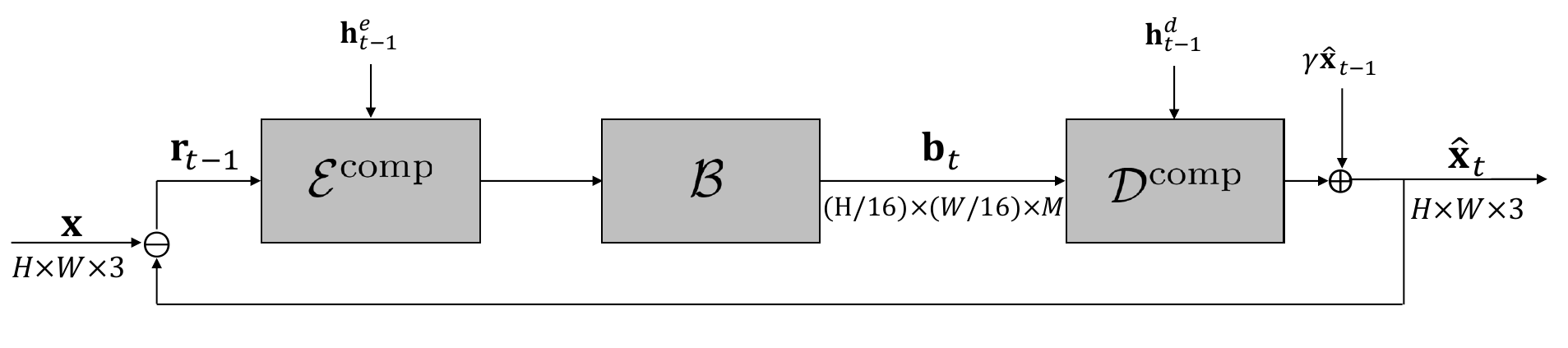}
     \caption{Compression RNN architecture. Here $\mathbf{x}$ is the original input image, $\mathbf{x}_{t}$ is the reconstructed (decoded) image after $t$ RNN iterations, $\mathbf{r}_{t-1}$ is the residual image from the previous iteration, $\mathbf{b}_{t}$ is the compressed binarized representation, and $\gamma \in \left\{ 0,1 \right\}$  regulates whether the reconstruction is additive or ``one shot". The architecture consists of a convolutional RNN based encoder $\mathcal{E}^{\rm{comp}}$ and decoder $\mathcal{D}^{\rm{comp}}$, whose hidden states are denoted by $\mathbf{h}_{t-1}^{e}$ and $\mathbf{h}_{t-1}^{d}$ respectively, and a stateless binarizer module $\mathcal{B}$.}
\label{fig:compression}
\end{figure*}

We now briefly discuss an RNN-based method for image compression proposed by \cite{toderici2017full}, which we use in our joint compression and 3D reconstruction framework.
The method is a single model architecture capable of producing variable rate compression (see Fig.~\ref{fig:compression}).
The encoder $\cal{E}^{\rm{comp}}$ and decoder $\cal{D}^{\rm{comp}}$ are RNN-based, therefore the processing occurs sequentially over time $t$.  
A single iteration is shown in Fig.~\ref{fig:compression}.
In the first iteration, the original image $\bf{x}$ is encoded by $\cal{E}^{\rm{comp}}$ to a vector of length $m$ in floating point representation, per element. Then, a binarizer module $\cal{B}$ converts the encoded representation into a binary vector. This binary vector is part of the compressed representation. In each further iteration, a new binary vector of length $m$ is formed and added to the full compressed representation. This way, the number of iterations controls the compression rate. 

Decoding applies $\cal{D}^{\rm{comp}}$ on the binary representation $\bf{b}_{t}$ and outputs either ($\gamma=0$) a full reconstructed image $\hat{\bf{x}}_{t}$ or ($\gamma=1$) a residual from the previous reconstruction $\hat{\bf{x}}_{t} - \hat{\bf{x}}_{t-1}$. At subsequent iterations, the input to the encoder becomes the residual image $\mathbf{r}_{t}=\mathbf{x}-\mathbf{x}_{t}$, which is the difference between the original and reconstructed image. During training, the absolute value of this residual is minimized.
The process is defined formally as follows
\begin{equation}
\label{eq:compression_1}
\hat{\mathbf{x}}_{t}={\cal{D}}_{t}^{\rm{comp}}\left(\mathbf{b}_{t}\right)+\gamma \hat{\mathbf{x}}_{t-1},~~~~
\mathbf{b}_{t}={\cal{B}}\left[ {\cal{E}}_{t}^{\rm{comp}} \left(\mathbf{r}_{t-1}\right)\right] ,\;\;\;\;\;
\end{equation}
\begin{equation}
\label{eq:compression_2}
\mathbf{r}_{t}=\mathbf{x}-\hat{\mathbf{x}}_{t},\;\;\;\; 
\mathbf{r}_{0} = \mathbf{x}, \;\;\;\;
\hat{\mathbf{x}}_{0}=0
\end{equation}
Here ${\cal{E}}_{t}^{\rm{comp}}$ and ${\cal{D}}_{t}^{\rm{comp}}$ represent the RNN-based encoder and decoder with their states at iteration $t$, respectively; $\mathbf{x}$ is an original image of size $H\times W \times 3$ and $\hat{\mathbf{x}}_{t}$ is its progressive reconstruction, with $\gamma=0$ for ``one-shot" reconstruction\footnote{Full image reconstruction is computed at each iteration}, and $\gamma=1$ for additive reconstruction\footnote{Only the residual is computed at each iteration. The final reconstructed image is a sum of the outputs of all iterations.}; $\mathbf{r}_{t}$ is the residual between $\mathbf{x}$ and the reconstruction $\hat{\mathbf{x}}_{t}$; and $\mathbf{b}_{t} \in \left \{ -1,1 \right \}^{m}$ is an encoded bit stream produced by a binarizer function $\cal{B}$,  where $m$ is the number of bits produced per iteration. It depends on $H$, $W$ and $M$, the number of output channels from the final layer of ${\cal{E}}^{\rm{comp}}$ through $m=\left(H/16\right)\times \left(W/16\right) \times M$ \footnote{$M=32$ in \cite{toderici2017full}.}. The number of RNN iterations $N$ controls the overall compression ratio. A compression rate $c$ is defined as the ratio of the number of bits in the raw image to that of the binarized representation. 
In our case we have 8 bit-per-pixel RGB images. Thus, in order to obtain a given compression rate $c$, the number of RNN iterations needs to be set to
\begin{equation}
\label{eq:compression_ratio}
N = \frac{H \times W \times 3 \times 8}{c\times m}.
\end{equation}
The encoder consists of a convolutional layer followed by three convolutional RNN layers, i.e. convolutional LSTMs. 
\cite{toderici2017full} evaluated additional options for the recurrence unit, besides LSTM.
The binarizer in \cite{toderici2017full} consists of a convolutional layer, followed by a binarizing operation, such as the one used in~\cite{toderici2015variable}. In \cite{toderici2017full}, further lossless compression is achieved using entropy coding.

During training, the binarizer in ~[\cite{toderici2015variable}, ~\cite{raiko2014techniques}] is a stochastic variable. 
Assuming the input $x$ is the output of a $\mathrm{tanh}$ layer, the binarizer is defined as
\begin{equation}
\label{eq:binarizer_2}
b(x)= 
\begin{cases}
1 & \textrm{with probability} ~~ \frac{1+x}{2}\\
-1 & \textrm{with probability} ~~ \frac{1-x}{2}
\end{cases}.
\end{equation}
For back-propagation, the derivative of the expectation is taken. Since, $\mathbb{E}\left [ b(x) \right ]=x ~~ \forall x\in \left [ -1,1 \right ]$, the gradients are passed through $b(x)$ unchanged. 

The decoder in \cite{toderici2017full} starts with a convolutional layer, followed by four convolutional RNN layers, with each such layer followed by a depth-to-space\footnote{The depth-to-space or Pixel Shuffle layer rearranges elements of a tensor of shape $\left(*,C \times k^{2},H,W\right)$ to a tensor of shape $\left(*,C, H\times k,W\times k\right)$. This, followed by a convolution layer is useful for upsampling sub-pixel convolution with a stride of $\frac{1}{k}$ } layer~(\cite{shi2016real}). Each such layer decreases the depth size by a factor of 4, thus increasing spatial resolution by a factor of 2 in both row and column dimensions. Finally, another convolutional layer is applied to produce a $H \times W \times3$ reconstructed image. 

During training, a weighted $L_{1}$ loss on the residual $
\mathbf{r}_{t}$ is minimized
\begin{equation}
\label{eq:compression_loss}
L_{\rm{comp}}=\beta \sum_{ijt=0}^{H W N}\left | {\bf{r}}_{ijt} \right | 
\end{equation}
where for minibatch size $B$, $\beta = \left( B \times H \times W \times 3 \times N \right) ^{-1}$ . The sum is over the image spatial dimensions $i,j$ and RNN iterations $t$.

\subsection{RNN-based 3D reconstruction}
\label{sec:background_3d_reconstruction}
\begin{figure*}
  \centering
    \includegraphics[width=0.8\textwidth]{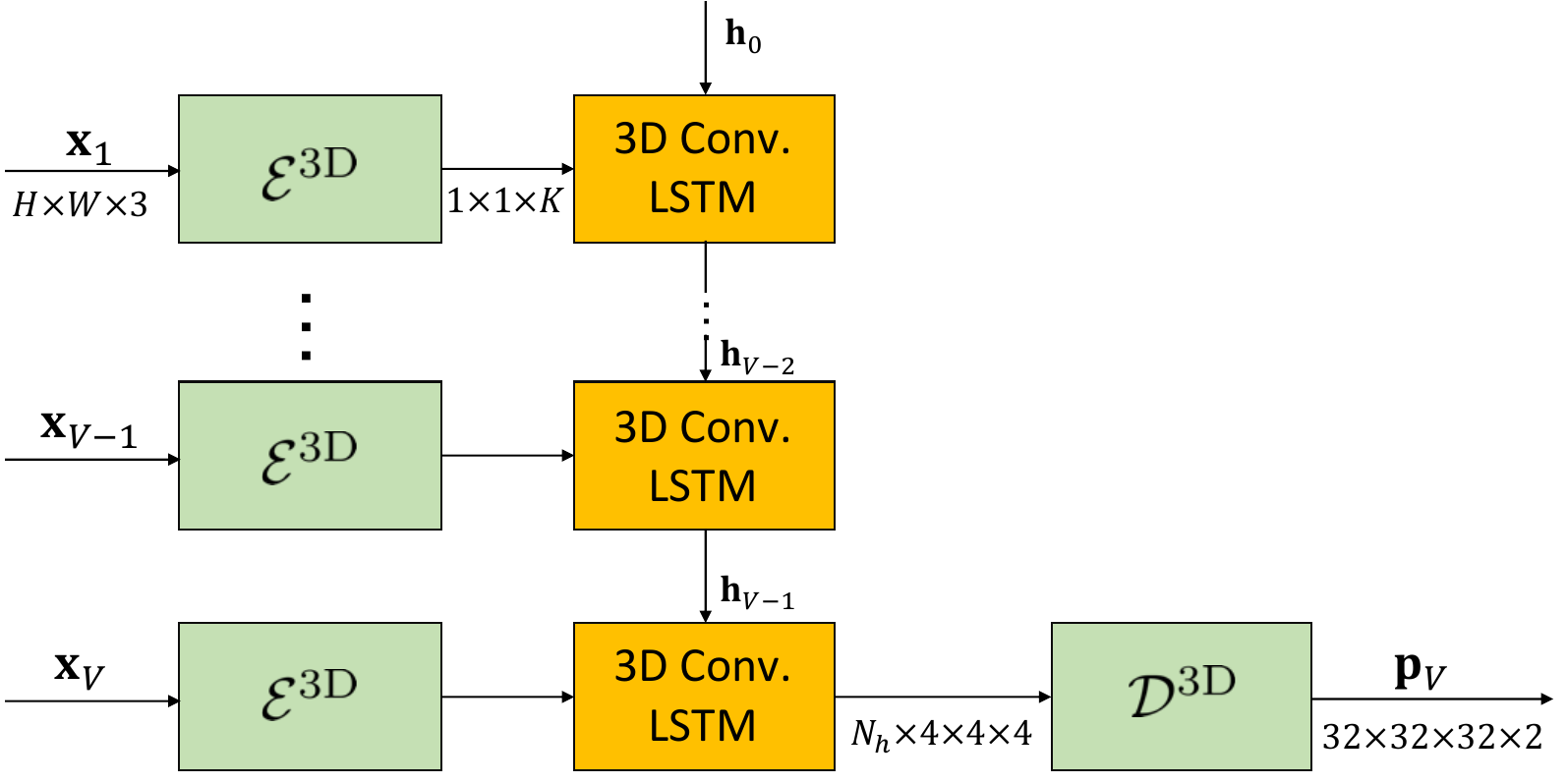}
     \caption{3D reconstruction network architecture. Here $\left\{\mathbf{x}_{i}\right\}_{i=1}^{V}$ are the input images, and $\mathbf{p}_{V}$ is the 3D reconstructed occupancy grid produced after $V$ viewpoints. The encoder ${\cal{E}}^{3D}$ produces a vector of size $K$ from each input image. These vectors enter a $4 \times 4 \times 4$  grid of 3D convolutional LSTM units, each with a hidden state $\mathbf{h}$ of size $N_{h}$. In the final view $V$, the hidden states are processed by a decoder ${\cal{D}}^{3D}$ to obtain the 3D reconstruction $\mathbf{p}_{V}$. Note that the same encoder ${\cal{E}}^{3D}$ and 3D convolutional LSTM units, with the same learned parameters are used for processing every viewpoint image.}
\label{fig:3d_reconstruction}
\end{figure*}

We now discuss the second important component for our joint framework for compression designed for 3D reconstruction, namely the 3D reconstruction method of \cite{choy20163d}. They proposed a network architecture whose input is a single or multiple images of an object, from arbitrary viewpoints. The network output is a reconstruction of the object in the form of a 3D occupancy grid (see Fig.~\ref{fig:3d_reconstruction}).
Here, $\left\{\mathbf{x}_{i}\right\}_{i=1}^{V}$ are images of an object from $V$ different viewpoints, and $\mathbf{p}_{V}$ is the 3D reconstructed occupancy grid produced after $V$ viewpoint. It is represented as Softmax probabilities indicating occupancy of each point in the 3D grid. 

The encoder $\cal{E}^{\rm{3D}}$ and decoder $\cal{D}^{\rm{3D}}$ are feedforward CNNs. Two architecture variants are proposed in \cite{choy20163d}: a shallow and a deeper one with residual blocks (\cite{he2016deep}).
Both encoder variants consist of convolutional layers followed by LeakyReLU nonlinearities, and six MaxPooling operations, each with a stride of 2, for a total encoder stride of 64. Finally, a fully-connected layer is applied, which returns a vector of size $K$, an embedding of each viewpoint image. 

The 3D convolutional LSTM module is a core component proposed in \cite{choy20163d}. It allows the network to retain details it has observed in previous views and update the memory when it receives a new image.
The module consists of a grid of $4 \times 4 \times 4$ 3D Convolutional LSTM units, each with a hidden state of size $N_{h}$. These are different from the convolutional LSTM units in Sec.~\ref{sec:background_compression} in two regards: First, the input of the LSTM module is a vector which undergoes multiplication as in a standard LSTM definition. Only the hidden state undergoes convolution. Second, with the hidden state being three-dimensional, the convolution operation is now in 3D. Each such unit is responsible for reconstructing a particular part of the 3D voxel space.
In \cite{choy20163d} the choices for $K$ and $N_{h}$ are 1024 and 128, respectively.

During training, a 3D voxel-wise Softmax loss over the final viewpoint's output $\mathbf{p}_{V}$, is minimized
\begin{equation}
\label{eq:3d_reconstruction_loss}
L_{\rm{3D}}= \sum \tilde{\mathbf{p}}\log \mathbf{p}_{V}+\left ( 1-\tilde{\mathbf{p}} \right )\log\left ( 1-\mathbf{p}_{V} \right ) 
\end{equation}
Here, $\tilde{\mathbf{p}}\in \left \{ 0,1 \right \}$ is the 3D ground-truth occupancy, and the sum is over the three voxel dimensions (indices 
omitted for simplicity).

The 3D Convolutional LSTM enables the network to be invariant to the order of viewpoints that are fed to it. Indeed, the $V$ viewpoints consisting each input sequence are selected at random, and there is no single preferred order. 
\section{Joint compression and 3D reconstruction}
\label{sec:compression_and_reconstruction}
The motivation for joining the tasks of compression and 3D reconstruction into a unified framework can be twofold. A primary motivation is to obtain compressed representations better suited for 3D reconstruction, and thus obtain improved 3D reconstruction performance, as compared to when using known compression standards.
Another motivation can be in reducing 
overall computational cost by providing a unified network architecture that would be more efficient than applying sequentially the compression and 3D reconstruction architectures shown in Figs.~\ref{fig:compression},~\ref{fig:3d_reconstruction}.

We now discuss different unified models that achieve the above needs.
In all of the following proposed network architectures, the loss is optimized with respect to all model parameters jointly. No sub-model elements are pre-trained, and no sequential training protocols take place, such as where one sub-model is trained first, then freezed, while another sub-model is trained. The Binarizer in all of the following proposed architectures has no learned parameters, but a gradient is passed through it as explained in Sec.~\ref{sec:background_compression}. 

\subsection{3D reconstruction from decoded images (Sequential model)}
\label{sec:3d_from_decoded_images}
\begin{figure*}
\begin{adjustwidth}{-0.15in}{-0.15in}
  \centering
    \includegraphics[width=1.05\textwidth]{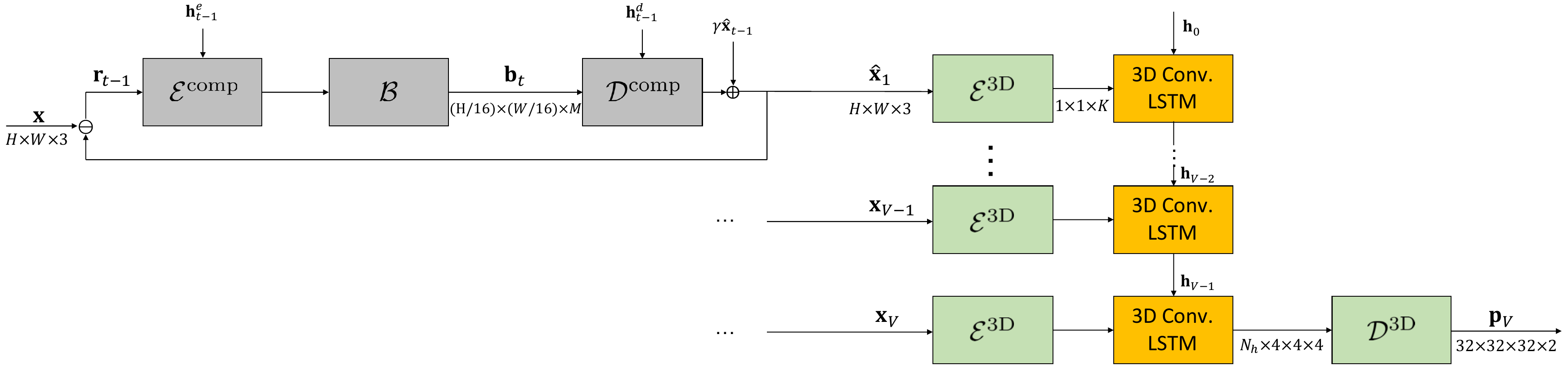}
     \caption{The \textit{sequential} model recovers a 3D occupancy grid from images decoded by a compression module. Note that the same compression encoder $\mathcal{E}^{\rm{comp}}$, binarizer $\mathcal{B}$, compression decoder $\mathcal{D}^{\rm{comp}}$, 3D reconstruction encoder $\mathcal{E}^{\rm{3D}}$, and 3D convolutional LSTM units, with the same learned parameters, are used for processing every viewpoint image.}
\label{fig:3d_from_decoded_images}
\end{adjustwidth}
\end{figure*}

Here we propose a sequential approach, as shown in Fig.~\ref{fig:3d_from_decoded_images}. The compression model (Fig.~\ref{fig:compression}) and 3D reconstruction model (Fig.~\ref{fig:3d_reconstruction}) are simply concatenated. The 3D reconstruction part of the network receives as input a decompressed image.
During training, we optimize a loss which depends on viewpoint $i$ as follows
\begin{equation}
\label{eq:total_loss}
L_{\rm{total}}= 
\begin{cases}
L_{\rm{comp}} & i<V\\
L_{\rm{comp}} + L_{\rm{3D}} & i=V
\end{cases}~.
\end{equation}
For $i<V$ we use $\cal{D}^{\rm{comp}}$ to obtain a decompressed image $\hat{\mathbf{x}}_{i}$, and the compression loss $L_{\rm{comp}}$ is applied and attempts to make $\hat{\mathbf{x}}_{i}$ similar to the original image $\mathbf{x}_{i}$.  In the last viewpoint $V$, this is done as well, but now, additionally the 3D reconstruction $\mathbf{p}_{V}$ is calculated, and the corresponding loss $L_{\rm{3D}}$ is also added to the optimization.
For simplicity, we refer to this model from now on as the \textit{sequential} model.
\subsection{3D reconstruction from compressed codes (Direct model)}
\label{sec:3d_from_compressed_codes}
\begin{figure*}
 \begin{adjustwidth}{-0.15in}{-0.15in}
  \centering
    \includegraphics[width=1.05\textwidth]{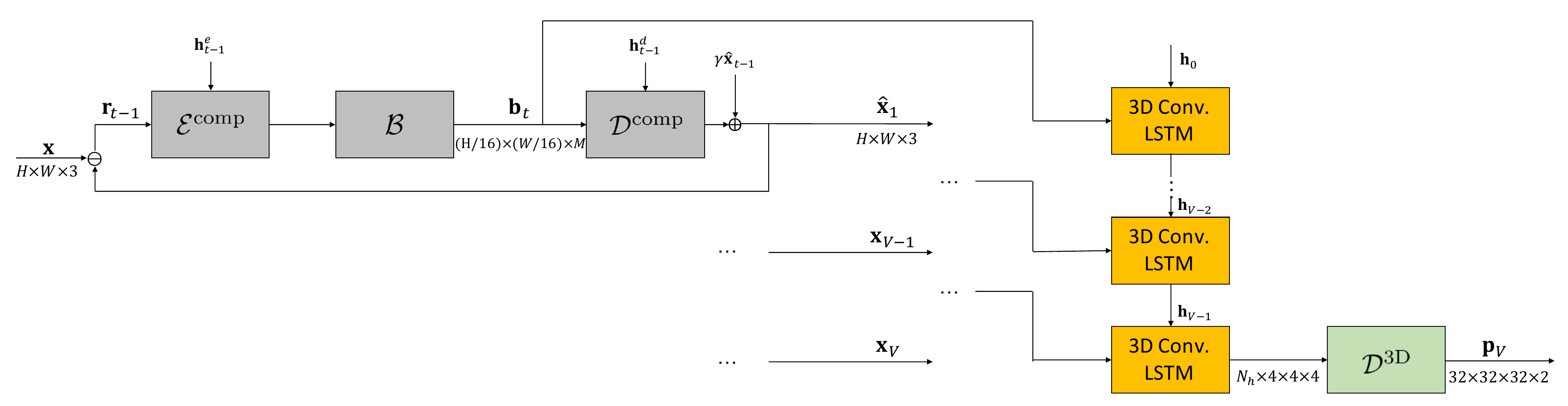}
     \caption{The \textit{direct} model recovers a 3D occupancy grid directly from the compressed codes. Note that the same compression encoder $\mathcal{E}^{\rm{comp}}$, binarizer $\mathcal{B}$, compression decoder $\mathcal{D}^{\rm{comp}}$, and 3D convolutional LSTM units, with the same learned parameters, are used for processing every viewpoint image.}
\label{fig:3d_from_compressed_codes}
\end{adjustwidth}
\end{figure*}

We also propose a more computationally efficient approach shown in Fig.~\ref{fig:3d_from_compressed_codes}. Here, the output codes following $\cal{E}^{\rm{comp}}$ and $\cal{B}$ are used directly to feed the 3D LSTM module. Thus, applying $\cal{E}^{\rm{3D}}$ is not required, and some computation is reduced.
Here we also optimize the loss from Eq.~(\ref{eq:total_loss}). 
For simplicity, we refer to this model from now on as the \textit{direct} model.
\subsection{3D reconstruction with implicit compression (Implicit model)}
\label{sec:3d_implicit_compression}
Finally, we propose another approach that is significantly more efficient computationally than those in Secs.~\ref{sec:3d_from_decoded_images}-\ref{sec:3d_from_compressed_codes}. 
The idea here is to use the RNN-based 3D reconstruction architecture discussed in Sec.~\ref{sec:background_3d_reconstruction}, and augment it with a binarizer $\cal{B}$ so that compression is obtained implicitly, without explicitly minimizing an image compression loss $L_{\rm{comp}}$. The loss we minimize here is simply $L_{\rm{3D}}$. This makes sense in a scenario where we wish to solve the 3D reconstruction task by supplying compressed image codes. We do not require that our model reconstruct viewable images. Only that it successfully solves the task at hand, which is to reconstruct a 3D occupancy map. Since we focus on the 3D reconstruction task, we can minimize $L_{\rm{3D}}$ only. 

The binarizer module $\cal{B}$ can be thought of as a form of regularization in training the network on the 3D reconstruction task. It imposes a constraint on the minimization of $L_{\rm{3D}}$, which is to yield a good 3D reconstruction while requiring the encoded representation to take binary form.
This architecture does not use the variable compression rate RNN framework of~\cite{toderici2017full}, where multiple RNN iterations controlled the compression rate. Rather, we train a separate slightly modified model for every desired compression rate. The compression rate is now controlled by $K$, the output vector length of the CNN encoder $\mathcal{E}_{K}^{\rm{3D}}$.

\begin{figure*}
  \centering
    \includegraphics[width=0.8\textwidth]{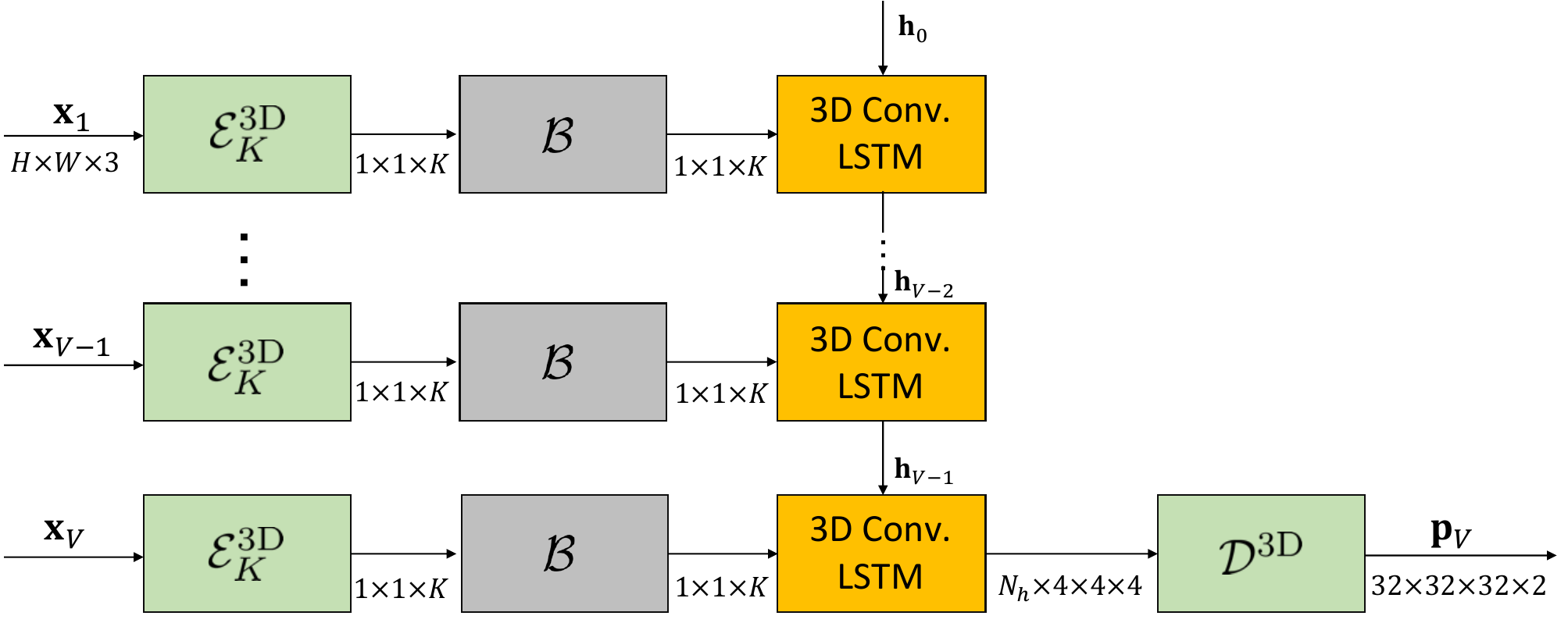}
     \caption{The \textit{implicit} model architecture recovers a 3D occupancy grid from compressed codes, without ever decompressing them into reconstructed images during training. Only a 3D reconstruction loss is minimized, thus image compression is obtained implicitly. Note that the same encoder $\mathcal{E}_{K}^{\rm{3D}}$, binarizer $\mathcal{B}$, and 3D convolutional LSTM units, with the same learned parameters, are used for processing every viewpoint image.}
\label{fig:3d_with_implicit_compression}
\end{figure*}

The proposed architecture is shown in Fig.~\ref{fig:3d_with_implicit_compression}. 
Here, the encoder $\mathcal{E}_{K}^{\rm{3D}}$ differs from $\cal{E}^{\rm{3D}}$ in that the default choice of $K$, the encoded vector length is now varied with every compression rate. A separate model is trained for different choices of $K$ obtained by modifying the number of channels in the output of the final convolutional layer of the encoder. Here $K$ is also the number of bits of the compressed representation (after the binarizer).
For simplicity, we refer to this model from now on as the \textit{implicit} model. 

Note that the \textit{direct} model from Sec.~\ref{sec:3d_from_compressed_codes} can be seen as a possible extension of the \textit{implicit} model, in case decompressed viewable images are required. It is similar in structure, except the encoder is $\mathcal{E}^{\rm{comp}}$ instead of $\mathcal{E}_{K}^{\rm{3D}}$, a decoder is added to produce the decompressed images, and decompression is done recurrently, allowing variable rate compression in a single architecture. 
\subsection{Implementation details}
\label{sec:implementation_details}
Our unified models were trained using a minibatch size of 6 so that training could be done on a single GTX 1080 Ti GPU. 
In Tab.~\ref{table:reimplementation_comparison} we compare our reimplementation of ~\cite{choy20163d} to the original work. Note that the model evaluated here is for 3D reconstruction only, without image compression involved. The evaluation criterion is mean Intersection-over-Union (mIoU) between a 3D voxel reconstruction $\mathbf{p}$ and its ground truth voxelized model $\mathbf{\tilde{p}}$, over the test set. The IoU criterion is defined [\cite{choy20163d}] as
\begin{equation}
\label{eq:iou}
\mathrm{IoU} \equiv \frac{\mathbbm{1}\left(\mathbf{p} > \tau\right) \cap \mathbbm{1}\left(\mathbf{\tilde{p}}\right)}{\mathbbm{1}\left(\mathbf{p} > \tau\right) \cup \mathbbm{1}\left(\mathbf{\tilde{p}}\right)} = \frac{\sum{\left[\mathbbm{1}\left(\mathbf{p} > \tau\right)\mathbbm{1}\left(\mathbf{\tilde{p}}\right)\right]}}  {\sum{\left[\mathbbm{1}\left(\mathbbm{1}\left( \mathbf{p} > \tau \right) + \mathbbm{1}\left(\mathbf{\tilde{p}}\right)\right)\right]}}~.
\end{equation}
Here, $\tau$ is a threshold we set to 0.4 as in \cite{choy20163d}, $\cap$ and $\cup$ denote the intersection and union operations, respectively, $\mathbbm{1}$ denotes the Indicator function, and the sums are over the three voxel dimensions.
Out of the different architecture variations that ~\cite{choy20163d} experimented with, we chose one of the simplest, termed 3D-LSTM-3. It consists of a network of moderate depth and uses an LSTM recurrence unit with a 3x3x3 convolution kernel.
\cite{choy20163d} trained the network for roughly 60 epochs. We see that we can train for only 20 epochs and still obtain good performance, only 3\% lower than the optimum. Therefore, for practical reasons, we settle on training all of our models for 20 epochs, from here onward.
We also see that our implementation results in somewhat improved performance compared to the original implementation of ~\cite{choy20163d}. This performance gain was not found to result from the few deviations from ~\cite{choy20163d} that are known to us, i.e, different minibatch size. We discuss the architecture of the compression part in our \textit{sequential} and \textit{direct} models (Secs.~\ref{sec:3d_from_decoded_images}-\ref{sec:3d_from_compressed_codes}) in the appendix.
\begin{table}[t]
\caption{Implementations of the 3D reconstruction method of ~\cite{choy20163d}. The evaluation criterion is mean IoU and standard deviation (STD) across the test set. (*) Here the error measure is an approximation of the STD. \cite{choy20163d} quote the mIoU $15\%$ and $85\%$ percentiles. Their difference is approximately $2\sigma$ assuming Gaussian distribution.  }
\label{table:reimplementation_comparison}
\begin{center}
\begin{tabular}{ll}
\multicolumn{1}{c}{\bf Implementation}  &\multicolumn{1}{c}{\bf mIoU}
\\ \hline \\
Choy et al. (original)         &$0.54 \pm 0.22$ (*)\\
Ours - 20 epochs             &$0.64 \pm 0.20$ \\
Ours - 60 epochs             &$0.66 \pm 0.20$ \\
\end{tabular}
\end{center}
\end{table}

In~\cite{toderici2017full}, the compression model was trained on small 32x32 image patches. In our \textit{sequential} and \textit{direct} proposed approaches (Secs.~\ref{sec:3d_from_decoded_images}-\ref{sec:3d_from_compressed_codes}), we want to build upon the architecture of ~\cite{toderici2017full} to train a unified model for both compression and 3D reconstruction. Therefore, we want the input to our models to be full resolution 128x128 images from the ShapeNet dataset used by~\cite{choy20163d}. Compared to ~\cite{choy20163d} and ~\cite{toderici2017full}, we wish to train larger models on larger inputs. We also wish to be able to do this on a single GPU. To handle this challenge, reducing the minibatch size to 6, as mentioned in Sec.~\ref{sec:implementation_details} was not enough. We needed to further optimize the compression network architecture. We provide elaborate discussion of our choice of compression network architecture in the appendix.

In this work we focus on the ShapeNet dataset of 128x128 rendered images. In different scenarios one may wish to apply our methods on much larger images. Our architectures are fully convolutional, therefore they can be directly be scaled to larger inputs with linear increase in memory and computation. However, the spatial receptive field would not grow unless further architectural changes are made, such as increasing the depth of the network, or using dilated convolutions (\cite{yu2015multi}).

\section{Results}
\label{sec:results}
\subsection{3D reconstruction}
\label{sec:results_3d_reconstruction}
\begin{figure*}[ht]
  \centering
    \includegraphics[width=0.8\textwidth]{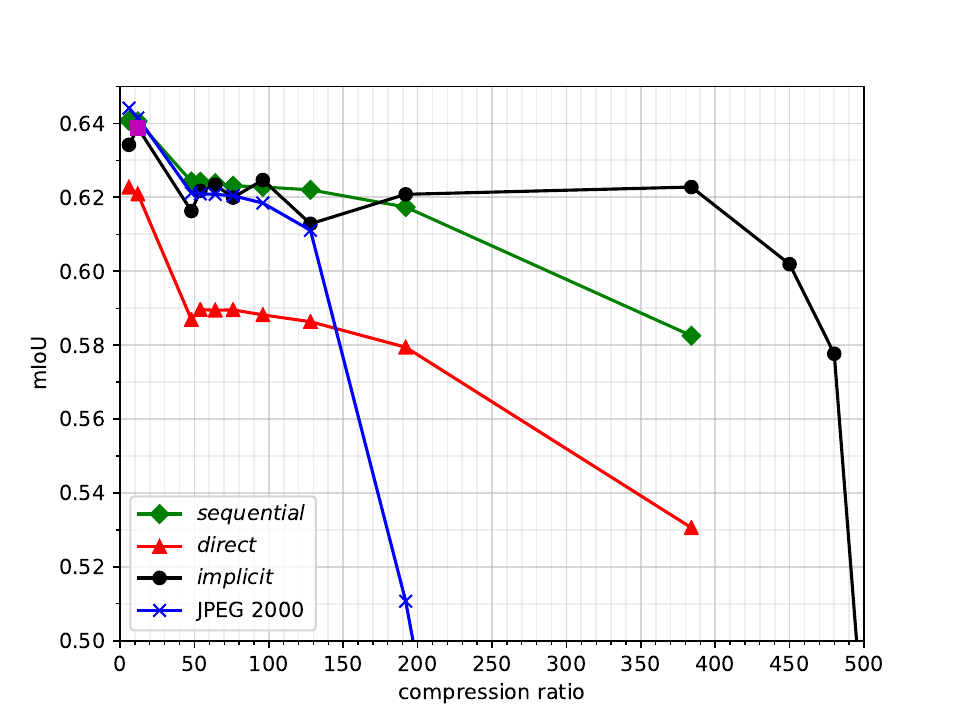}
     \caption{3D reconstruction mIoU scores on the ShapeNet test set, using 5 viewpoints. We compare a network as in \cite{choy20163d} trained and applied on JPEG 2000 compressed images (blue), to our \textit{sequential} (green), \textit{direct} (red) and \textit{implicit} (black) models. The magenta square at a compression rate of 12 denotes our reimplementation of the original \cite{choy20163d} model. The standard deviation (STD) across the test set for all models and compression rates is approximately 0.2.}
\label{fig:results}
\end{figure*}

In Fig.~\ref{fig:results}, we report our models' mean Intersection-over-Union (mIoU) scores on the ShapeNet test set, using 5 viewpoints for the 3D reconstruction, as in ~\cite{choy20163d}. Tab.~\ref{table:unified_models_visual_3d_comparison} shows a visual comparison of 3D reconstructed occupancy grids produced for 5 random viewpoints of the objects shown in Tabs.~\ref{table:unified_models_visual_compression_comparison},~\ref{table:visual_compression_arch_comparison}.

\begin{table*}[ht]
\caption{Visual comparison of our models' 3D reconstruction capability, for different compression rates. The occupancy grids are produced for the same four objects that were used in Tabs.~\ref{table:unified_models_visual_compression_comparison},~\ref{table:visual_compression_arch_comparison}. The 3D reconstruction for uncompressed images is shown identically in both rows for convenience.}
\centering
\label{table:unified_models_visual_3d_comparison}
\begin{adjustwidth}{-0.25in}{-0.5in}
\begin{tabular}{|m{0.16\textwidth}|m{0.16\textwidth}|m{0.16\textwidth}|m{0.16\textwidth}|m{0.16\textwidth}|m{0.1\textwidth}|}
\hline\hfill
\begin{center}
\textbf{1:1 (uncompressed)}\end{center}& \hfill\begin{center}
\textbf{48:1}\end{center} & \hfill\begin{center}\textbf{192:1} \end{center}&\hfill \begin{center}\textbf{384:1} \end{center}&\hfill \begin{center}\textbf{480:1} \end{center} &\\\hline
                          \includegraphics[width=1\linewidth]{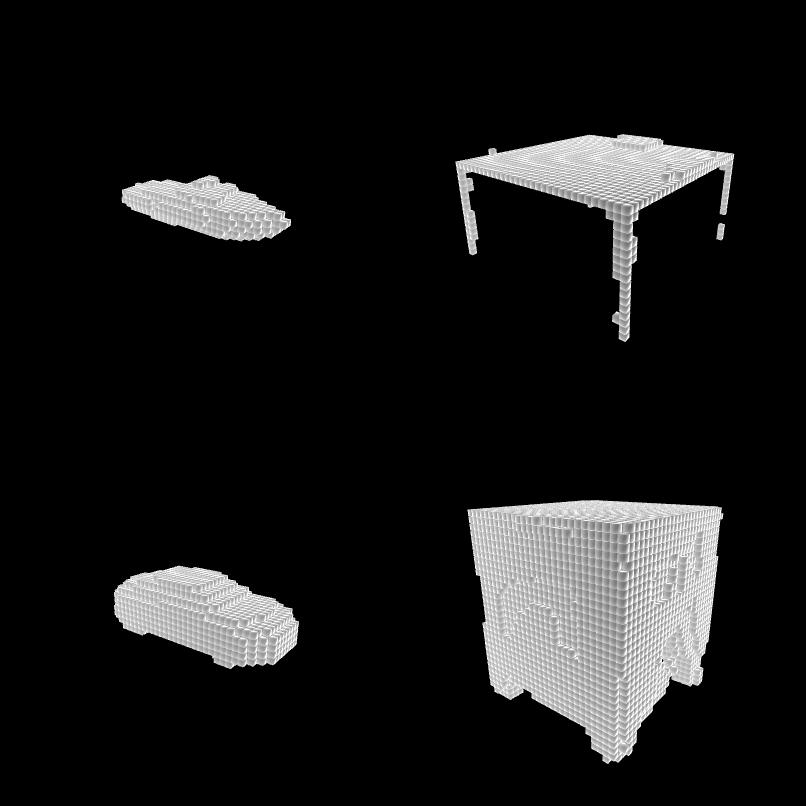}&             \includegraphics[width=1\linewidth]{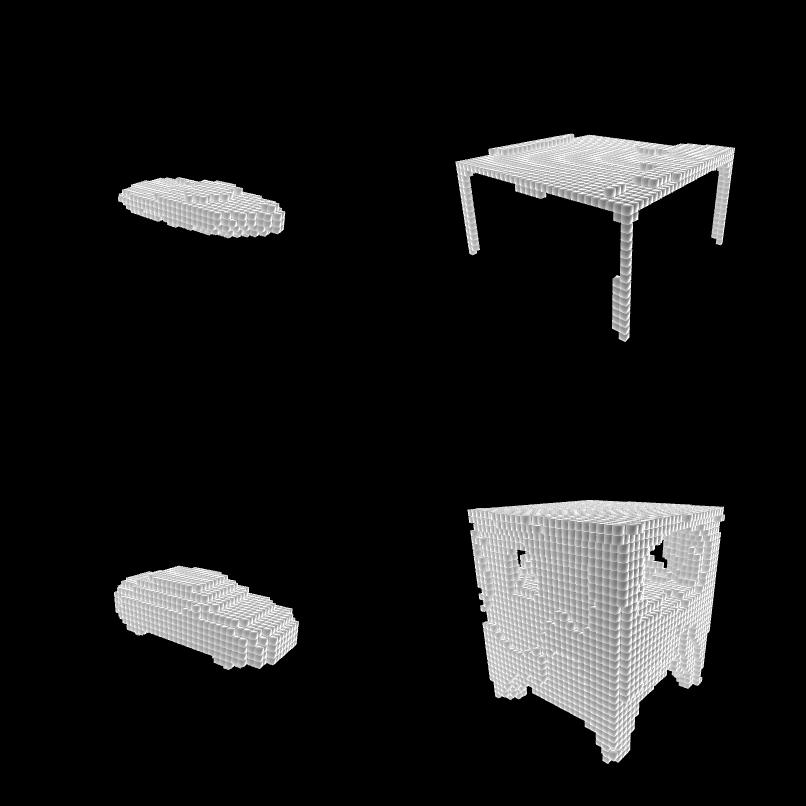}&              \includegraphics[width=1\linewidth]{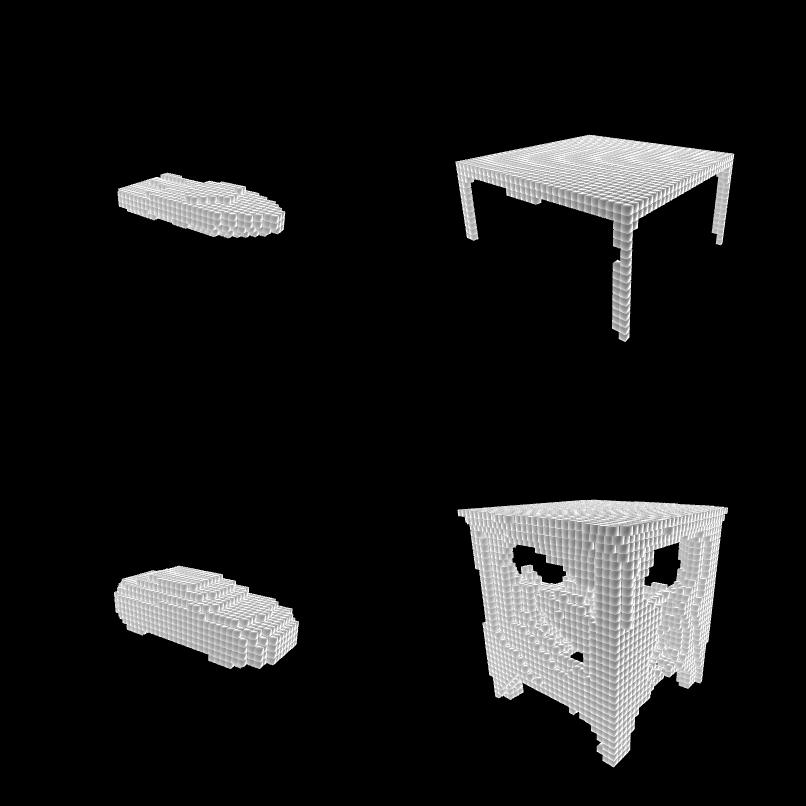}&              \includegraphics[width=1\linewidth]{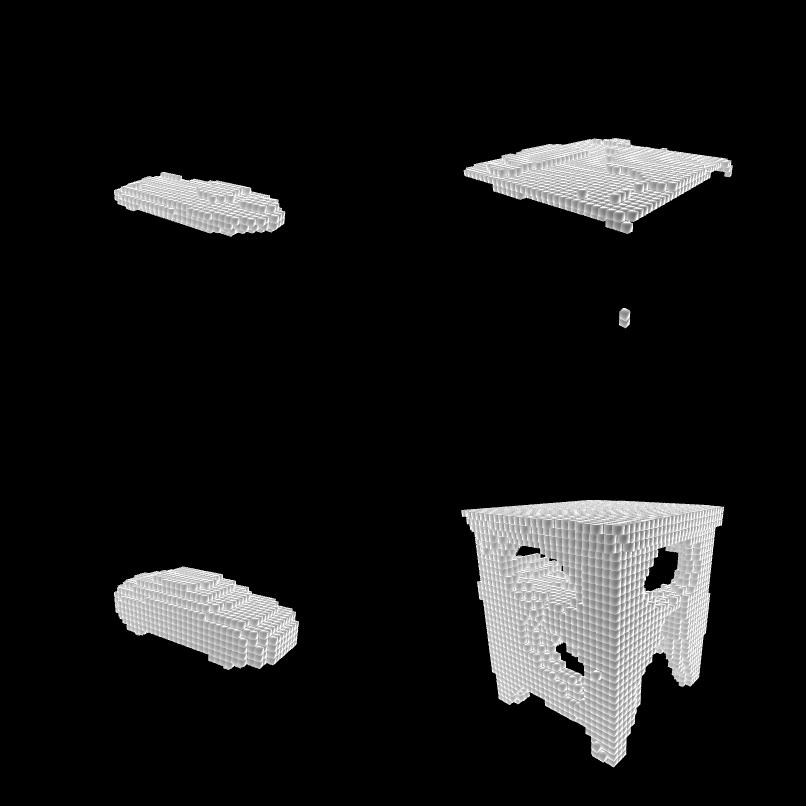}& ~& \makecell{\textbf{\textit{Sequential} }\\ \textbf{model}\\  } \\ \hline
                          \includegraphics[width=1\linewidth]{fig_images/3d_visual_comparison/uncompressed.jpg}&             \includegraphics[width=1\linewidth]{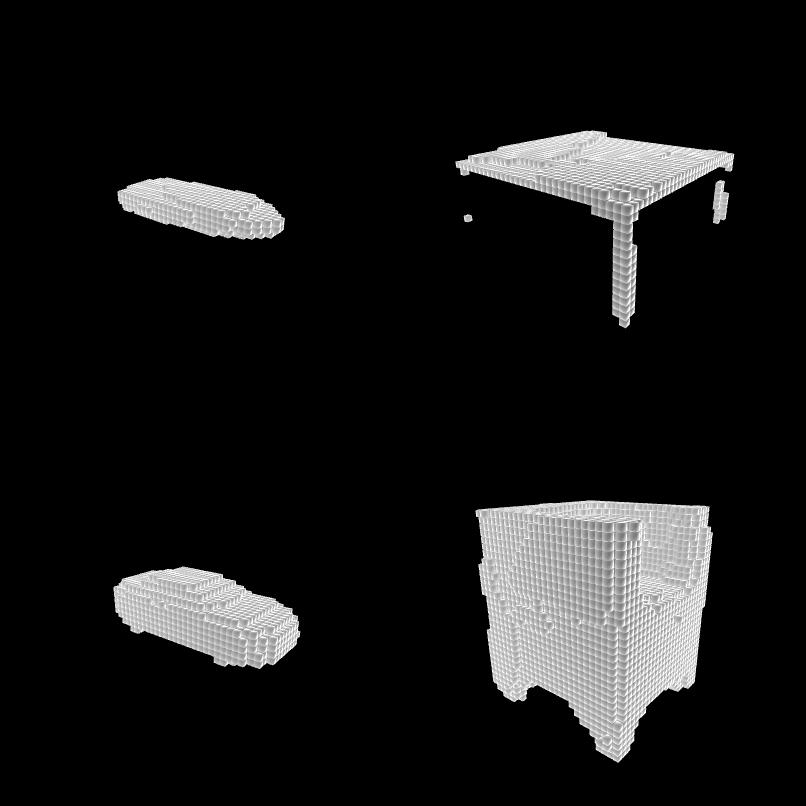}&              \includegraphics[width=1\linewidth]{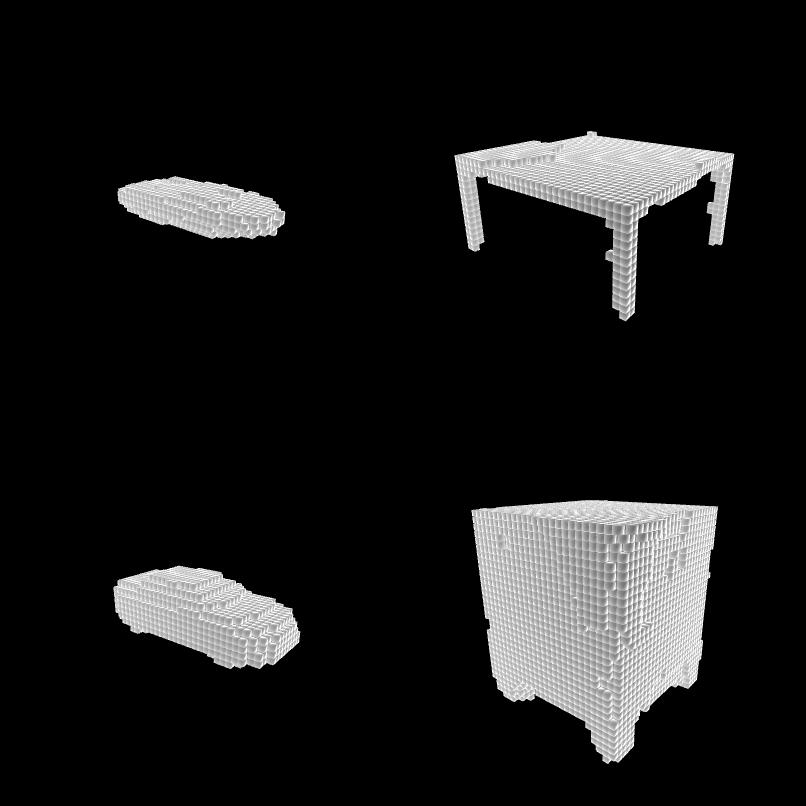}&              \includegraphics[width=1\linewidth]{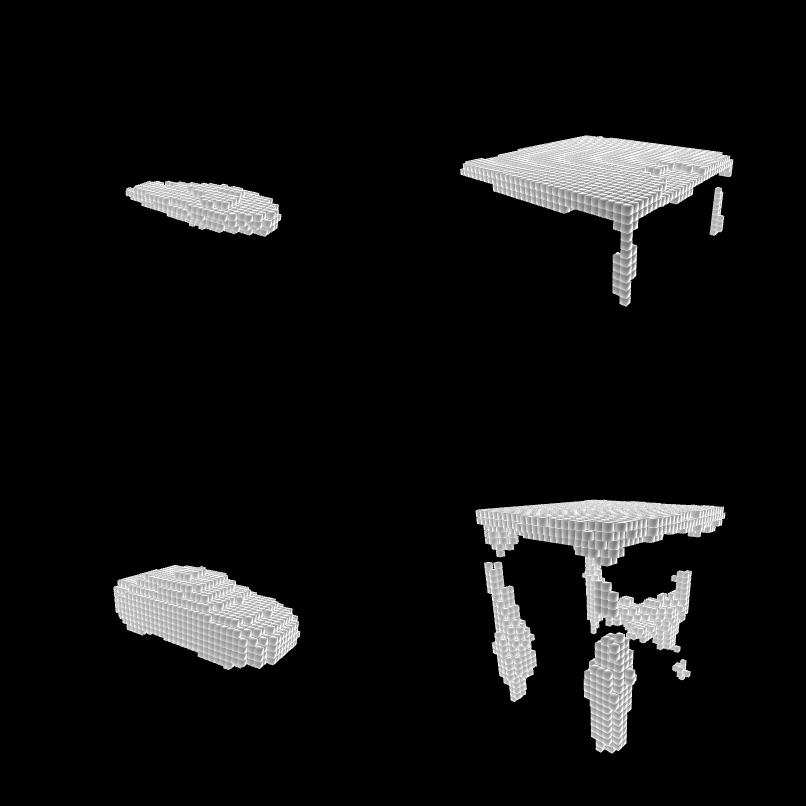}& ~ &\makecell{\textbf{\textit{Direct} }\\ \textbf{model}\\} \\ \hline
\includegraphics[width=1\linewidth]{fig_images/3d_visual_comparison/uncompressed.jpg}&            \includegraphics[width=1\linewidth]{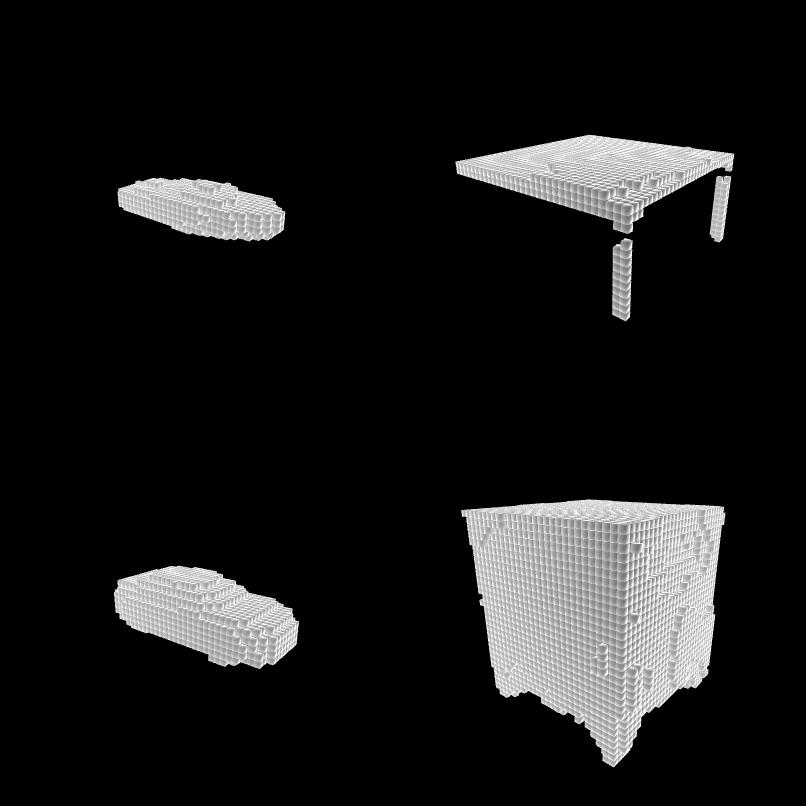}&            \includegraphics[width=1\linewidth]{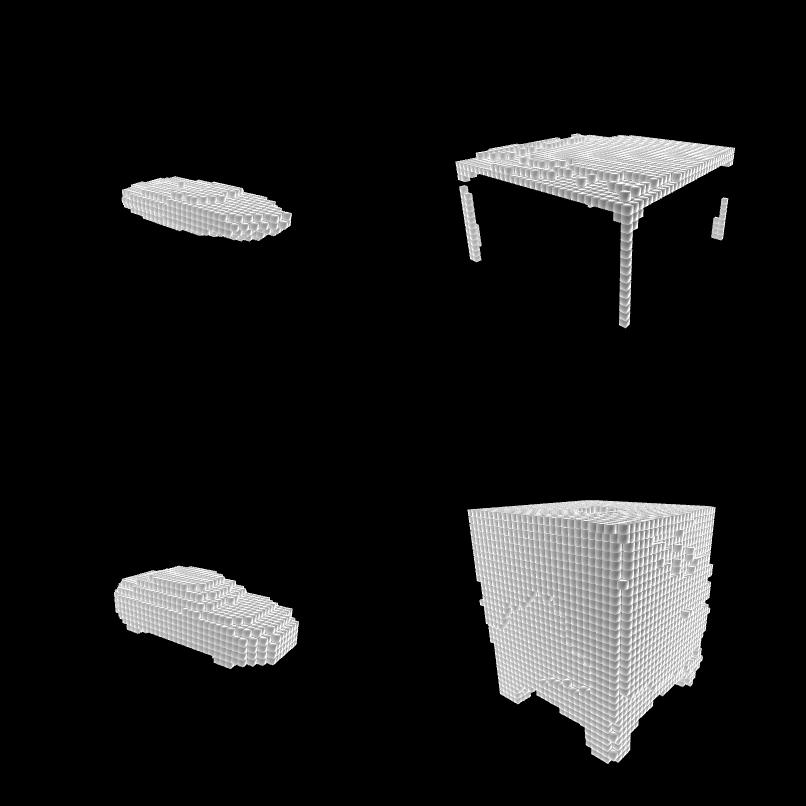}&            \includegraphics[width=1\linewidth]{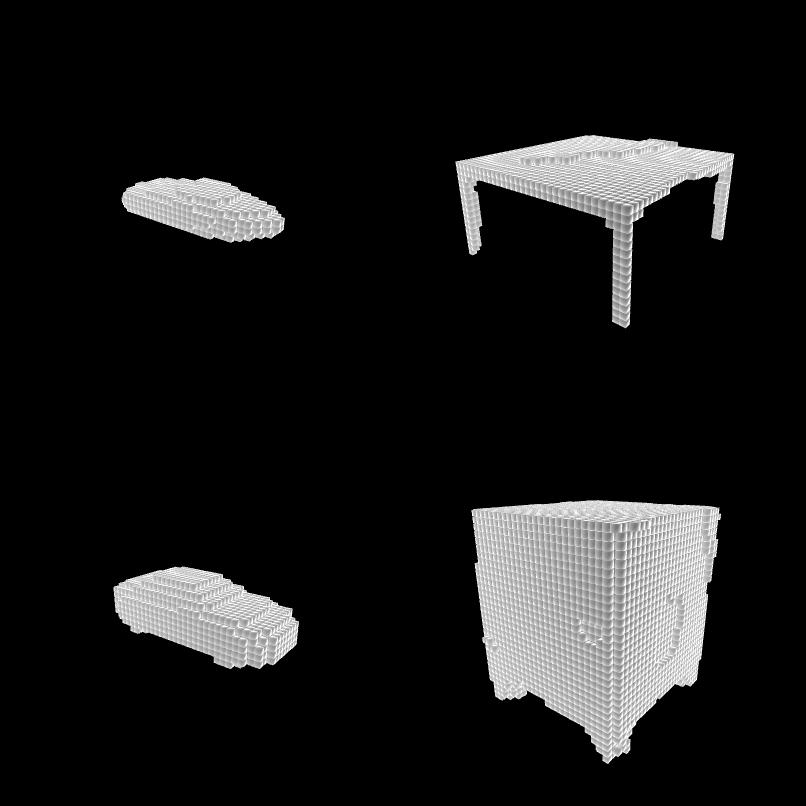}&  \includegraphics[width=1\linewidth]{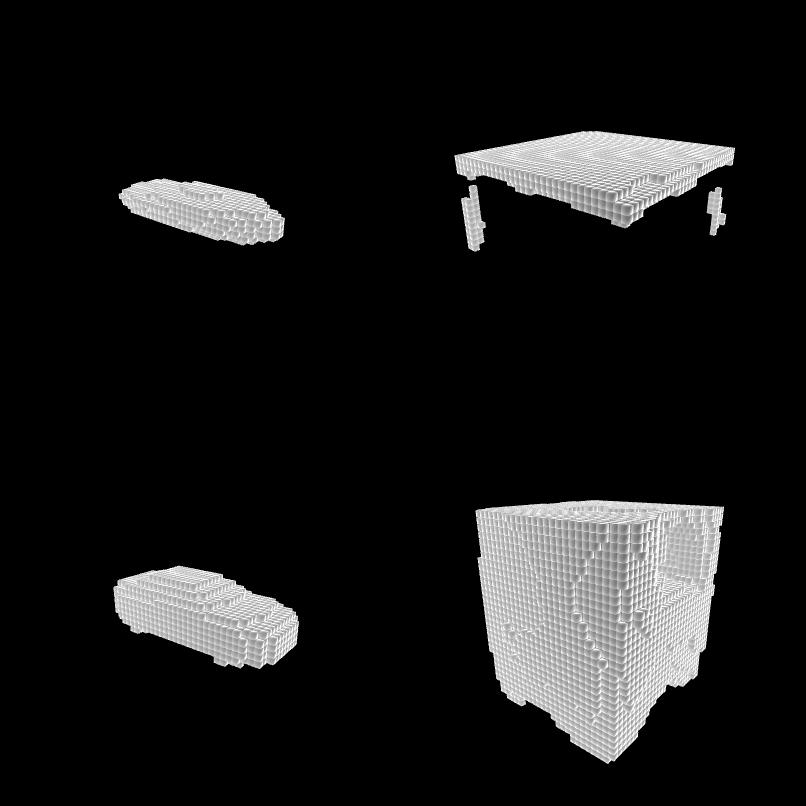} & \makecell{\textbf{\textit{Implicit}}\\ \textbf{model}\\ }           \\ \hline                          
\end{tabular}
\end{adjustwidth}
\end{table*} 

We see that for aggressive compression rates of 128 and above, the \textit{sequential} model is superior to a 3D reconstruction network trained on images compressed by JPEG 2000 using a random compression rate at each minibatch. Moreover, for an extremely aggressive compression rate of 384, JPEG 2000 typically compresses images to the extent that no details are seen at all. The mIoU of the network trained on JPEG 2000 compressed images at this point is 0.1, which is meaningless. In our \textit{sequential} and \textit{direct} models, this compression ratio is obtained by using only one RNN timestep. It still produces images that contribute to 3D reconstruction with $\rm{mIoU}>0.5$. We also see that the best results are obtained using the most efficient \textit{implicit} model. 
\subsection{Statistical significance}
\label{sec:statistical_significance}
\begin{figure}
  \centering
    \includegraphics[width=0.5\textwidth]{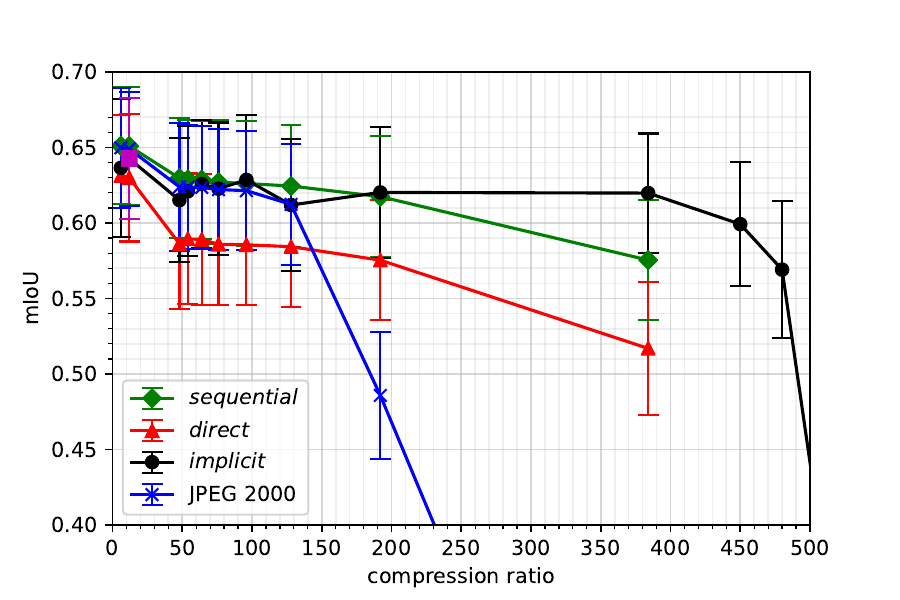}
     \caption{3D reconstruction mIoU scores shown with error bars, on a subset of 340 examples from the ShapeNet test set, for which the STD is 0.04 (on the baseline model of \cite{choy20163d}), using 5 viewpoints. We compare a network as in \cite{choy20163d} trained and applied on JPEG 2000 compressed images (blue), to our \textit{sequential} (green), \textit{direct} (red) and \textit{implicit} (black) models. The magenta square at a compression rate of 12 denotes our reimplementation of the original \cite{choy20163d} model.}
\label{fig:results_5th_chunk}
\end{figure}

The 3D reconstuction IOU metric varies significantly for different test examples, even without compression. This is due to the natural variability between objects. Different objects can be easier or more challenging for accurate 3D reconstruction than others. For the baseline 3D reconstruction model of ~\cite{choy20163d}, the STD of the IOU across the test examples is roughly 0.2 (see Tab.~\ref{table:reimplementation_comparison}). We ask ourselves then, how significant are some of the trends depicted in Fig.~\ref{fig:results}? To obtain a better insight for this, we separate the test examples into bins of roughly equal IOU STD of 0.04 on the baseline model. For this value of STD, we obtain 7 such bins which contain (in increasing order of mean IOU) 17, 82, 131, 273, 340, 384 and 233 examples. To illustrate this, we show the mIOU results for bin \#5 with error bars, in Fig.~\ref{fig:results_5th_chunk}. Here, we see more clearly that the performance trends and differences between models, are of statistical significance.
In the two lowest mIOU data bins, the mIOU for all the different models and most compression rates is below 0.2. For such low IOU examples, the evidence for statistical significance is lacking. It may also be explained by the relatively small number of examples that are contained in these low IOU bins. 
\subsection{Lower compression rates}
\label{sec:results_lower_compression_rates}
Note that the original network of \cite{choy20163d} naturally obtains an image compression ratio of 1:12, without any use of a binarizer (the magenta square in Fig.~\ref{fig:results}) or otherwise special design. It can be thought of as a special case of the \textit{implicit} model, with the binarizer module disabled, and set to $K=1024$ in $\mathcal{E}_{K}^{\rm{3D}}$. Thus, the magenta square is plotted on top of the \textit{implicit} model curve.

This work focused on high, limiting compression rates. Still, we wanted to provide a more complete overview of our methods. Therefore, we additionally evaluated all our models for two lower compression rates of 1:12 and 1:6. We can say roughly that all models tend to approach the ``ideal" case of the magenta square at these low compression rates. 

If we used our \textit{sequential}, \textit{direct} and \textit{implicit} architectures with binarizer, obtaining such low compression rates would become a memory burden. Therefore, we disabled the binarizer and used a 32 bit floating point encoded vector. We set $N$ the number of RNN iterations, or the number of channels in the last encoder layer $K$ (for the \textit{implicit} architecture) appropriately in order to achieve these rates. 
For the network trained on JPEG 2000 compressed images, we re-trained it separately for 1:12 and 1:6 compression rates. 
\subsection{Run time}
\label{sec:run_time}
Tab.~\ref{tab:timing} shows time of execution for different proposed models and building blocks on a single GTX 1080 Ti GPU, during both train (forward and backward pass) and test time (forward pass only)\tablefootnote{The backward pass time of $\mathcal{E^{\rm{3D}}}$ is approximated. It was not measured directly since it was done through the whole 3D reconstruction module with a single PyTorch .backward() call. It is reasonable to assume that it is identical to the forward pass, as this was the case when measured on the whole 3D reconstruction module.
The timing of the joint 3D reconstruction with implicit compression model (Sec.~\ref{sec:3d_implicit_compression}) is approximated as identical to just the 3D reconstruction model of~\cite{choy20163d}, since it only adds a single binarizer element on top of it, that is negligible in computational complexity compared to the whole model in both forward and backward passes.}.
A typical scenario of 5 viewpoints and 4 RNN iterations was assumed when timing 3D reconstruction and compression modules, respectively. 
We see that the using $\mathcal{N^{\rm{small}}}$ as the compression module in our \textit{sequential} unified model results in a speed-up of 15\% in the forward pass compared to using $\mathcal{N^{\rm{original}}}$. Using the more efficient \textit{direct} model further improves the relative speed-up percentage to 26\%. 

This improvement is moderate, implying that the most of the computation lies within the recurrent compression module. One way to increase the speed-up could be in eliminating the variable compression rate constraint and train a model per compression rate. This is done in the \textit{implicit} model, however without reconstructing viewable decompressed images.

The most significant speed-up is not surprisingly obtained by the \textit{implicit} model which only optimizes $L_{\rm{3D}}$, and simply augments the 3D reconstruction model by~\cite{choy20163d} with a binarizer module $\cal{B}$ (Sec.~\ref{sec:3d_implicit_compression}). Here, the relative speed-up percentage in forward pass reaches 75\%. Note that we trained all models on the same data and number of epochs, so these relative speed-ups also indicate the difference between whole training procedures required to obtain a deployable model.  

\begin{table*}[]
\caption[]{Execution time of different models and building blocks. 3D reconstruction was measured for 5 viewpoints and compression was done using 4 RNN iterations. Bold numbers in parentheses denote relative time improvement percentage compared to sequentially applying compression as in~\cite{toderici2017full} followed by 3D reconstruction as in~\cite{choy20163d}}
\label{tab:timing}
\begin{center}
\begin{tabular}{lll}
\textbf{Model / Block} & \textbf{Forward pass [msec]} & \textbf{Backward pass [msec]} \\ \hline
3D Reconstruction            & 13             & 13              \\
3D Reconstruction ($\cal{E}^{\rm{3D}}$ only)            & 6             & $\approx$~6 \\
Compression - $\cal{N}^{\rm{original}}$.~\small{(\cite{toderici2017full}) }           & 40             & 24 \\  
Compression - $\cal{N}^{\rm{small}}$. (See appendix)            & 32             & 20 \\
\textit{Sequential} model (Sec.~\ref{sec:3d_from_decoded_images})            & 45 \bf{(15\%)}            & 33 \bf{(11\%)}\\
\textit{Direct} model (Sec.~\ref{sec:3d_from_compressed_codes})            & 39 \bf{(26\%)}            & 27 \bf{(27\%)}\\
\textit{Implicit} model (Sec.~\ref{sec:3d_implicit_compression})            & $\approx$ ~13 \bf{(75\%)}            & $\approx$ ~13 \bf{(65\%)}\\  
\end{tabular}
\end{center}
\end{table*}

\subsection{Compressed image decoding}
\label{sec:results_compression}
\begin{figure}
  \centering
    \includegraphics[width=0.5\textwidth]{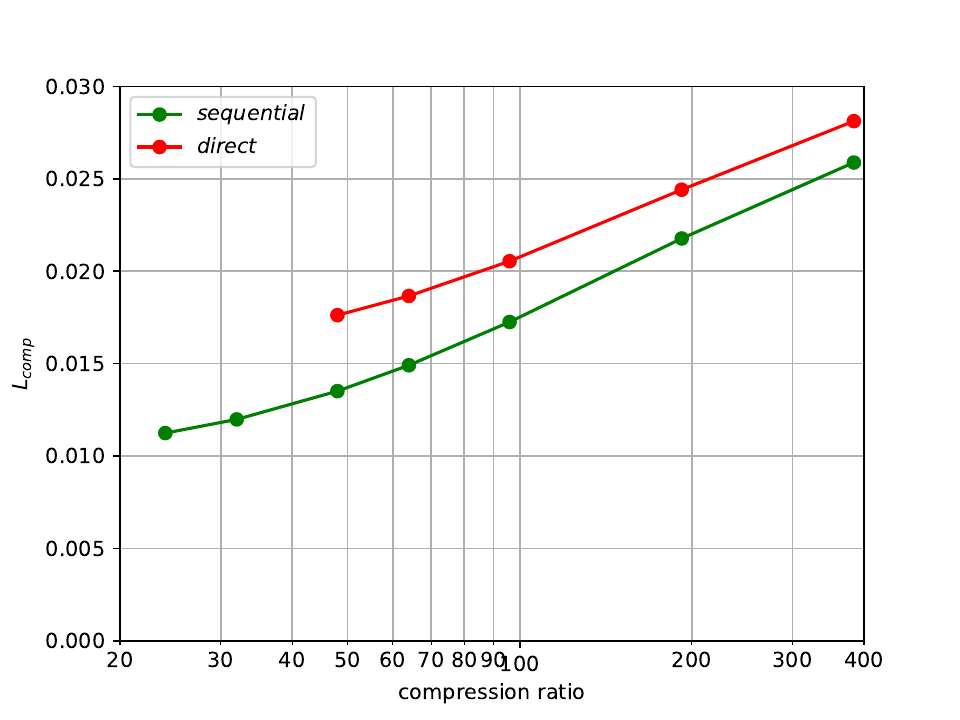}
     \caption{Compression loss $L_{\rm{comp}}$ obtained on the ShapeNet test set by the \textit{sequential} and \textit{direct} model architectures.}
\label{fig:results_compression}
\end{figure}

Although our goal is 3D reconstruction from compressed image representations, it is useful to also be able to decompress into visually satisfying 2D images. Our \textit{sequential} and \textit{direct} models allow this, since they were also optimized for the 2D image decoding task. Here, we evaluate their performance on this task.
Fig.~\ref{fig:results_compression} shows the compression loss magnitude that our models achieve on the ShapeNet test set, and Tab.~\ref{table:unified_models_visual_compression_comparison} shows a visual comparison.
We see that the compressed image decoding performance of the \textit{sequential} model is virtually indistinguishable from that of $\mathcal{N^{\rm{small}}}$, a model trained on the image compression task only (See appendix). In contrast, the more computationally efficient \textit{direct} model is somewhat inferior.
This could be explained by a harder constraint that this model imposes on the compression codes, requiring them to be suitable for directly feeding the 3D reconstruction model, in addition to allowing for a good compressed image decoding.

\begin{table*}[]
\begin{adjustwidth}{-0.25in}{-0.35in}
\caption{Visual comparison of the \textit{sequential} and \textit{direct} models when probed for their compression capability, for different compression rates. Four different images from the ShapeNet test set are shown. The uncompressed images are shown identically in both rows for convenience.}
\centering
\label{table:unified_models_visual_compression_comparison}
\begin{tabular}{|m{0.21\textwidth}|m{0.21\textwidth}|m{0.21\textwidth}|m{0.21\textwidth}|m{0.1\textwidth}|}
\hline\hfill
\begin{center}\textbf{1:1 (uncompressed)}\end{center} & \hfill\begin{center}\textbf{48:1}\end{center} & \hfill\begin{center}\textbf{192:1}\end{center} & \hfill\begin{center}\textbf{384:1}\end{center} &                    \\ \hline
                          \includegraphics[width=1\linewidth]{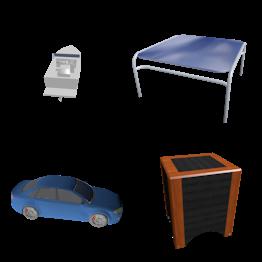}&             \includegraphics[width=1\linewidth]{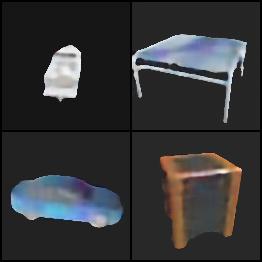}&              \includegraphics[width=1\linewidth]{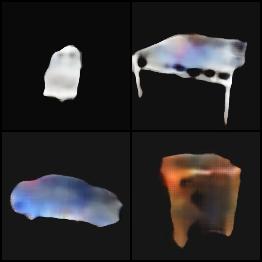}&              \includegraphics[width=1\linewidth]{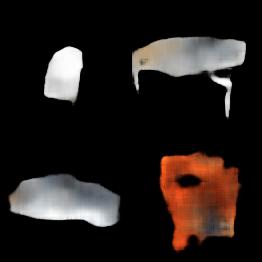}& \makecell{\textbf{\textit{Sequential} }\\ \textbf{model} } \\ \hline
                          \includegraphics[width=1\linewidth]{fig_images/joint_visual_comparison/uncompressed.jpg}&             \includegraphics[width=1\linewidth]{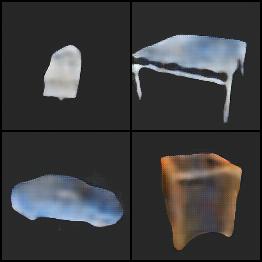}&              \includegraphics[width=1\linewidth]{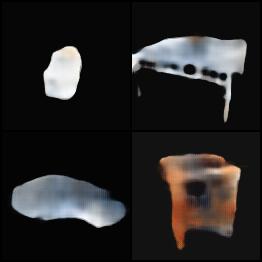}&              \includegraphics[width=1\linewidth]{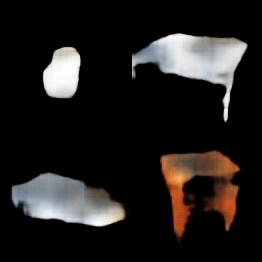}& \makecell{\textbf{\textit{Direct} }\\ \textbf{model} } \\ \hline
\end{tabular}
\end{adjustwidth}
\end{table*} 

In order to gain more qualitative insight on the decompression capability, we show in Tab.~\ref{table:more_texture_visual_comparison} another visual example, with four other images which were manually selected to contain richer texture content. Here we also show JPEG-2000 decompression results for comparison.
We can see that for high compression rates, the neural network based compression is superior to JPEG-2000, however in the 48:1 rate, JPEG-2000 manages to keep details better, particularly in areas of rich, high spatial frequency texture. This may show that such image details were not essential to learn good 3D reconstruction, in our case where the 3D models ground truth is relatively low resolution (32x32x32 voxel grids).
\begin{table*}[]
\begin{adjustwidth}{-0.25in}{-0.35in}
\caption{Visual comparison of the \textit{sequential} and \textit{direct} models when probed for their compression capability, and JPEG-2000, for different compression rates. Four different images from the ShapeNet test set are shown. Relatively detailed texture images were manually selected. The uncompressed images are shown identically in both rows for convenience.}
\centering
\label{table:more_texture_visual_comparison}
\begin{tabular}{|m{0.21\textwidth}|m{0.21\textwidth}|m{0.21\textwidth}|m{0.21\textwidth}|m{0.1\textwidth}|}
\hline\hfill
\begin{center}\textbf{1:1 (uncompressed)}\end{center} & \hfill\begin{center}\textbf{48:1}\end{center} & \hfill\begin{center}\textbf{192:1}\end{center} & \hfill\begin{center}\textbf{384:1}\end{center} &                    \\ \hline
                          \includegraphics[width=1\linewidth]{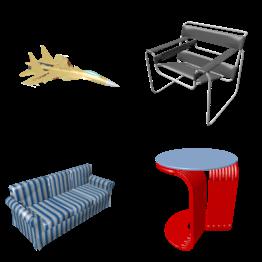}&             \includegraphics[width=1\linewidth]{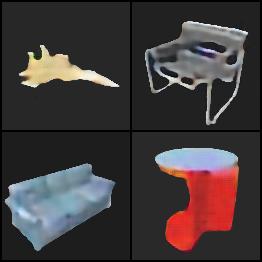}&              \includegraphics[width=1\linewidth]{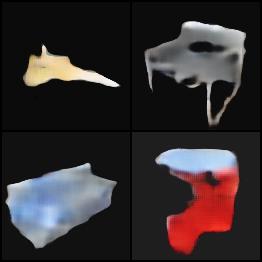}&              \includegraphics[width=1\linewidth]{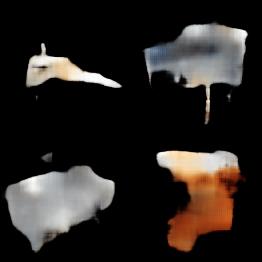}& \makecell{\textbf{\textit{Sequential} }\\ \textbf{model} } \\ \hline
                          \includegraphics[width=1\linewidth]{fig_images/more_texture_visual_comparison/uncompressed.jpg}&             \includegraphics[width=1\linewidth]{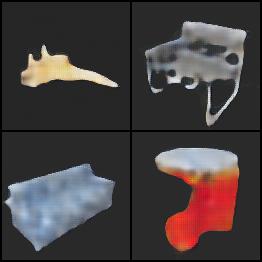}&              \includegraphics[width=1\linewidth]{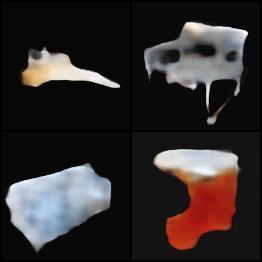}&              \includegraphics[width=1\linewidth]{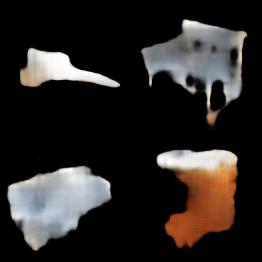}& \makecell{\textbf{\textit{Direct} }\\ \textbf{model} } \\ \hline
\hline
                          \includegraphics[width=1\linewidth]{fig_images/more_texture_visual_comparison/uncompressed.jpg}&             \includegraphics[width=1\linewidth]{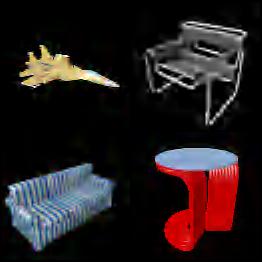}&              \includegraphics[width=1\linewidth]{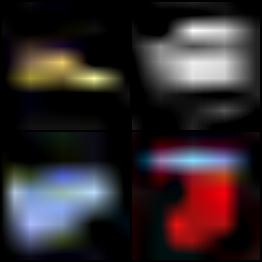}&              \includegraphics[width=1\linewidth]{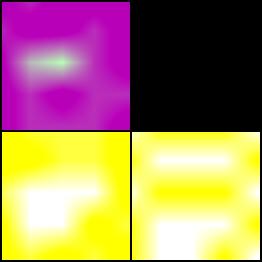}& \makecell{\textbf{\textit{JPEG} }\\ \textbf{2000} } \\ \hline            
\end{tabular}
\end{adjustwidth}
\end{table*} 
\section{Discussion}
\label{sec:discussion}
\vspace{-\topsep}
The main conclusions from the paper can be summarized as follows:
\vspace{-\topsep}
\begin{itemize}
\itemsep0em
  \item Three neural network architectures for joint image compression and 3D reconstruction were proposed which trade-off 3D reconstruction quality, image decompression, and run time.
  \item Near optimal 3D reconstruction is obtained for a wide range of compression rates, including aggressive rates where JPEG-2000 fails completely.
  \item The \textit{implicit} architecture reaches the most aggressive compression rates without performance loss, while also being the fastest in run time.
\end{itemize}
\vspace{-\topsep}
Let us now discuss in more detail. The \textit{sequential} and \textit{direct} model architectures include an RNN-based module dedicated for image compression and allows for compressed image decoding. These architectures are trained to optimize two loss functions 
simultaneously, one for 2D image compression, and another for 3D reconstruction. As such, they imply significant additional computations (both in train and test time) compared to the 3D reconstruction module alone. The \textit{implicit} model architecture is both more computationally efficient and accurate. It is trained only by minimizing the loss function associated with the 3D reconstruction task. The compression is obtained implicitly via a binarizer module which is inserted as part of the 3D reconstruction network architecture. The computational overhead that this variant imposes on top of an existing 3D reconstruction module alone, is negligible.

In future work, techniques for speeding up training time of current models, especially the \textit{sequential} and \textit{direct}, could be explored. Training speed-up could be obtained by not training from scratch as we did, and instead using pre-trained weights. Training speed-up could also be obtained by using more efficient base models for compression and 3D reconstruction. Finally, existing numerical techniques such as mixed precision training could be applied.

We showed that all our proposed architectures allow for reasonable 3D reconstruction from images compressed aggressively at a 
compression ratio of 384:1, where JPEG-2000 compression is impractical. Two of our more accurate architectures also outperform RNN-based 3D reconstruction trained on JPEG-2000 compressed images at less 
aggressive compression rates. Our best performing \textit{implicit} architecture outperforms RNN-based 3D reconstruction trained on JPEG-2000 compressed images for almost the whole range of compression rates, while also reaching acceptable 3D reconstruction performance under even more aggressive compression rates than all other examined variants (up to 480:1). 

The \textit{implicit} model yields relatively constant and nearly ``ideal", non-compressed performance throughout a wide range of compression rates. This motivated us to try apply it using even more aggressive compression rates then the previously chosen 384. And indeed we see that reasonable results are still obtained for up to a compression rate of 480, after which they deteriorate quickly.

The \textit{implicit} model is not optimized for acceptable image recovery. It only considers 3D reconstruction. This has limitations, in that such a model could only be used in specific applications, where a human observer is not present, or is not interested in the images, but only in the final 3D reconstructed occupancy grids. However, this also means that this model solves a much simpler problem. It does not need to account for two constraints simultaneously. Thus, for suitable applications, it is highly beneficial. It achieves superior 3D reconstruction capability over a very broad range of compression rates. And it does so using much less computations. 

The somewhat noisy, non-monotonic trend of the mIoU vs. compression rate of the \textit{implicit} model may perhaps be explained by that in contrast to the \textit{sequential} and \textit{direct} models, here, a separate model is trained for every compression rate, so it is reasonable that there can be some degree of inconsistency. 

Another reason could be related to our choice of varying the compression rate by modifying the number of output channels from the last layer of ${\mathcal{E}}^{3D}_{K}$. This choice may not necessarily be the best one.
One could modify the architecture in different ways to control the compression rate, and it may be that restricting the modification to just a single layer is somewhat sub-optimal and may result in some overfitting if the number of neurons in a single layer is too large. This could explain the slightly lower performance counter-intuitively obtained for better compression rates. 

Some techniques can be suggested in order to obtain a more smooth performance curve across compression rates.
Increasing regularization through weight decay, dropout or other known techniques, could lead to more stability. Additionally, it's possible to use a common backbone architecture with identical weights across all compression rates, while only retraining the last convolutional layer of the encoder, for each rate. This could also reduce variance across compression rates. 

We also note that for low compression rates, not using the binarizer and setting $K$ accordingly is one possible design choice, and not necessarily the most optimal one. We could, for instance, use a quantizer that outputs 4, 8 or 16 bits, together with a different choice of $K$. Or modify the encoder architecture in another way. We leave the focus on moderate to low compression rates for future work.

We find it interesting that 3D reconstruction based on neural networks can be made highly robust to image compression, with only slight performance degradation for an extremely wide range of compression rates. We believe that it may be worth exploring neural networks' robustness to significant image compression for other computer vision tasks as well.

Another promising direction for future work may be to further improve the achievable compression rates, by considering the overlap between adjacent image views in the compression module, instead of compressing the images separately before feeding the compressed representations to the 3D reconstruction module.

We note again that our approach is meant to propose a generic concept which offers benefit in jointly framing the tasks of image compression and 3D reconstruction. Both these separate tasks are well researched and the state-of-the-art in each of them has been improved in recent years over the methods we chose to use in our models.

\newpage
\begin{appendices}
\section*{Appendix - Choice of compression architecture}
Due to the challenge in single GPU training of network architectures that extend RNN-based compression modules, on full resolution 128x128 images, we propose a \textit{smaller} compression network architecture $\mathcal{N^{\rm{small}}}$ relative to the \textit{original} one $\mathcal{N^{\rm{original}}}$ proposed in ~\cite{toderici2017full}. 
We now elaborate the differences between them. 

In our proposed smaller architecture, the 3rd (last) convolutional RNN layer of the encoder $\mathcal{E^{\rm{comp}}}$ outputs 16 feature maps instead of 512. Therefore, the encoded representation is already compact and the binarizer does not include a convolutional layer. The binarized representation vector length in $\mathcal{N^{\rm{small}}}$ is 16, rather than 32 in $\mathcal{N^{\rm{original}}}$. This allows us to achieve a more aggressive maximal compression ratio of 384 vs. 192 (The maximal compression rate is obtained when just a single compression RNN iteration is used. Lower rates are obtained as desired using more iterations). 
The decoder $\mathcal{D^{\rm{comp}}}$ in our proposed $\mathcal{N^{\rm{small}}}$ consists of only three convolutional RNN layers instead of four, and they are smaller in capacity compared to $\mathcal{N^{\rm{original}}}$.

The 3D reconstruction model in ~\cite{choy20163d} used a random number of viewpoints during training. This was useful for obtaining a model that at test time, can reconstruct shape from an arbitrary number of viewpoints. In contrast, ~\cite{toderici2017full} trained their image compression model using a constant, maximal number of RNN iterations, which corresponded to a compression ratio of 24:1. We propose to train $\mathcal{N^{\rm{small}}}$ and our models based on it, as described in Sec.~\ref{sec:compression_and_reconstruction}, using a number of RNN iterations (and thus, compression rate), selected at random in each training iteration. This significantly reduces training time, and also facilitates slightly more robust performance across varying compression rates. 

\begin{figure}
  \centering
    \includegraphics[width=0.5\textwidth]{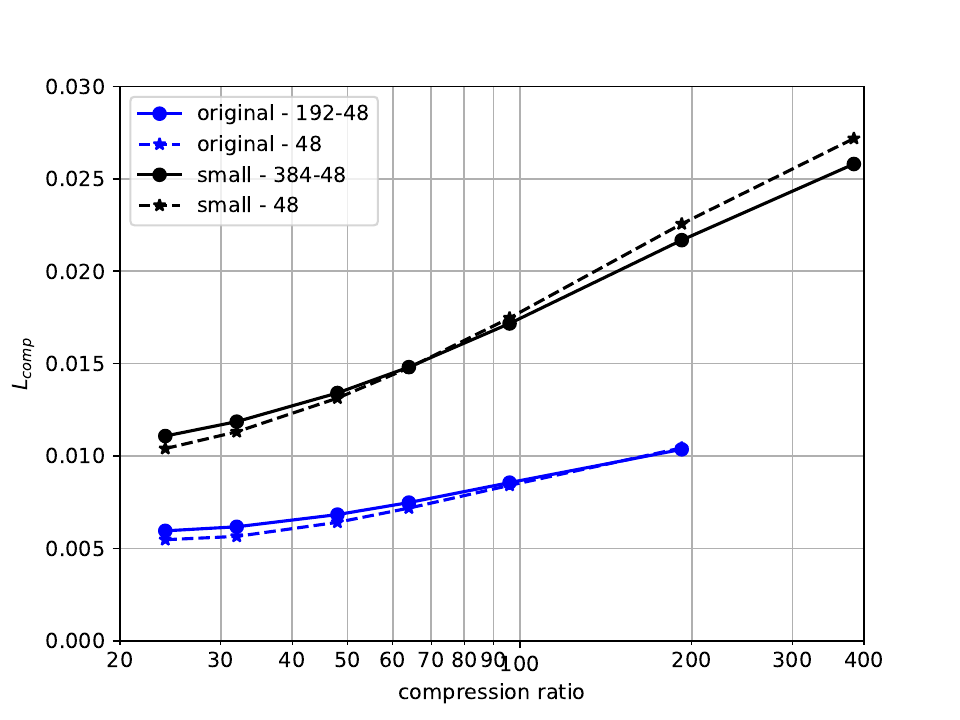}
     \caption{$L_{\rm{comp}}$ for $\mathcal{N^{\rm{small}}}$ and $\mathcal{N^{\rm{original}}}$. Each alternative was trained using a constant number of RNN iterations corresponding to a compression rate of 1:48, and randomly varying up to a rate of 1:384.}
\label{fig:compression_arch_quantitative_comparison}
\end{figure}

Fig.~\ref{fig:compression_arch_quantitative_comparison} compares the loss $L_{\rm{comp}}$ over the ShapeNet test set of $\mathcal{N^{\rm{small}}}$ to that of $\mathcal{N^{\rm{original}}}$, using both constant and randomly varying number of RNN iterations. We see a significant gap of around a factor of 2 in terms of the loss value in favor of $\cal{N}^{\rm{original}}$. 
To gain an intuition of the visual effect of such gap, we show in Tab.~\ref{table:visual_compression_arch_comparison} a comparison of a selection of four images from the ShapeNet test set, decoded using the different architectures, at different compression rates. We see a noticeable difference in quality in favor of $\cal{N}^{\rm{original}}$, relative to $\cal{N}^{\rm{small}}$. 
Despite the difference in visual image quality, we use $\cal{N}^{\rm{small}}$ throughout our experiments. In Sec.~\ref{sec:results}, we show that this choice is still appropriate given that our actual task of interest is that of 3D reconstruction. On this task, the performance that our unified model achieves is only a few percent lower than that achieved using uncompressed images, for a wide range of compression rates.
 
\begin{table*}[]
\caption{Visual comparison of different compression methods for different compression rates. Four different images from the ShapeNet test set are shown. The uncompressed images are shown identically in all three rows for convenience.}
\centering
\label{table:visual_compression_arch_comparison}
\begin{tabular}{|c|c|c|c|c|}
\hline
\textbf{1:1 (uncompressed)} &\textbf{48:1} & \textbf{192:1} & \textbf{384:1} &                    \\ \hline
                          \includegraphics[width=0.2\linewidth]{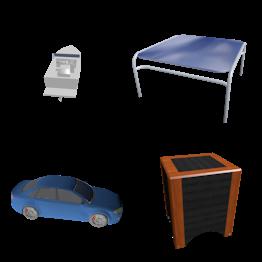}&             \includegraphics[width=0.2\linewidth]{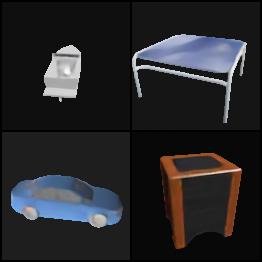}&              \includegraphics[width=0.2\linewidth]{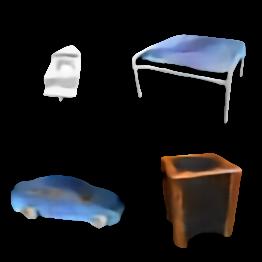}&              & \makecell{$\cal{N}^{\mathbf{original}}$ } \\ \hline
                          \includegraphics[width=0.2\linewidth]{fig_images/visual_comparison/uncompressed.jpg}&             \includegraphics[width=0.2\linewidth]{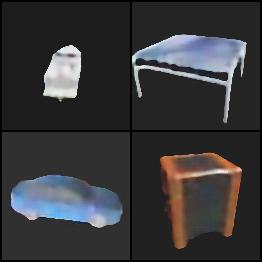}&              \includegraphics[width=0.2\linewidth]{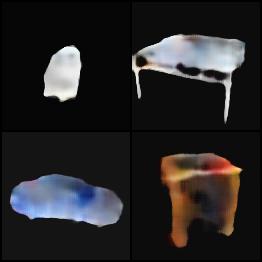}&              \includegraphics[width=0.2\linewidth]{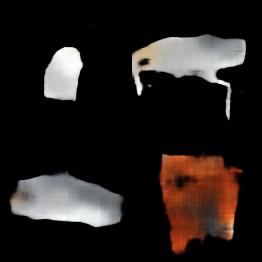}& \makecell{\textbf{$\cal{N}^{\mathbf{small}}$ }} \\ \hline
                          \includegraphics[width=0.2\linewidth]{fig_images/visual_comparison/uncompressed.jpg}&             \includegraphics[width=0.2\linewidth]{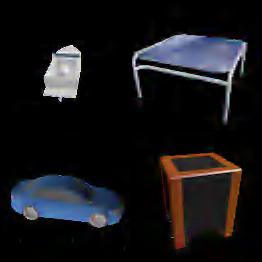}&              \includegraphics[width=0.2\linewidth]{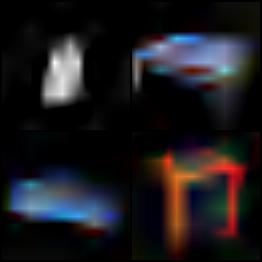}&              \includegraphics[width=0.2\linewidth]{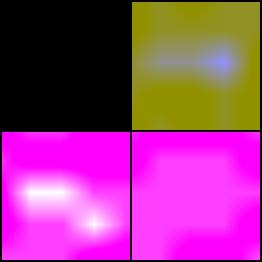}& \textbf{JPEG-2000} \\ \hline
\end{tabular}
\end{table*} 
\end{appendices}
\section*{Acknowledgments}
We would like to thank Christopher Choy for his responsiveness to our questions about his work, as well as Alona Golts for her helpful insights.

Yoav Schechner is the Mark and Diane Seiden Chair in Science at the Technion. He is a Landau Fellow - supported by the Taub Foundation. His work is conducted in the Ollendorff Minerva Center. Minvera is funded through the BMBF. This project is funded by the European Research Council (ERC) under the European Union’s Horizon 2020 research and innovation program (grant agreement No 810370: CloudCT) and the Technion Autonomous Systems Program (TASP).

\bibliographystyle{model1-num-names.bst}
\bibliography{yjvci-revised}

\end{document}